\documentclass[10pt,journal,compsoc]{IEEEtran}
\pdfoutput=1

\usepackage{booktabs}
\usepackage{balance}
\usepackage[mathscr]{euscript}
\usepackage{verbatim}
\usepackage{amssymb,amsmath,color,graphicx, amsbsy, bm}
\usepackage{subcaption}
\usepackage{epsfig}
\usepackage{amsthm}

\usepackage{fixltx2e}
\usepackage{xcolor}

\allowdisplaybreaks[1]

\usepackage{bbm}
\usepackage{amssymb}
\usepackage{algorithmic}       
\usepackage{amsmath}
\usepackage{graphicx}
\usepackage{booktabs}
\usepackage[flushleft]{threeparttable}
\usepackage{multirow}
\usepackage{amsthm}
\usepackage{graphicx,booktabs,multirow}
\usepackage{stmaryrd}
\usepackage{multirow,url}
\usepackage{algorithm}
\usepackage{cite}
\usepackage{color}

\newtheorem*{assumption*}{Assumption}
\ifodd 1

\newcommand{\com}[1]{\textbf{\color{red} (Comment: #1) }}
\newcommand{\comg}[1]{\textbf{\color{blue} (COMMENT: #1)}}
\newcommand{\response}[1]{\textbf{\color{blue} (RESPONSE: #1)}}
\else

\newcommand{\com}[1]{}
\newcommand{\comg}[1]{}
\newcommand{\response}[1]{}
\fi

\def\T{\mathcal{T}}

\def\I{\mathcal{I}}
\def\N{\mathcal{N}}
\def\M{\mathcal{M}}

\def\s{\boldsymbol{s}}
\def\a{\boldsymbol{a}}
\def\Ss{\mathcal{S}}
\def\As{\mathcal{A}}

\def\thetae{\boldsymbol{\theta}^{\text{Eval}}}
\def\thetat{\boldsymbol{\theta}^{\text{Target}}}
\def\Qe{{Q}^{\text{Eval}}}
\def\Qt{{Q}^{\text{Target}}}
\def\Qtmp{\hat{\boldsymbol{Q}}^{\text{Target}}}
\def\qtmp{\hat{Q}^{\text{Target}}}
\def\Atmp{\boldsymbol{a}^{\text{Next}}}
\def\actset{\mathcal{B}}
\def\actnum{B}

\def\arriv{\lambda}
\def\arrivedge{\arriv^{\text{edge}}}

\def\density{\rho}
\def\iotsev{f^{\text{device}}}

\def\trans{f^{\text{tran}}}

\def\hiscon{\boldsymbol{H}}

\def\ts{T^{\text{step}}}

\def\delay{d}

\def\Bfog{q^{\text{edge}}}
\def\boldBfog{\boldsymbol{q}^{\text{edge}}}
\def\fogsev{f^{\text{edge}}}

\def\task{k}

\def\ttran{{l}^{\text{tran}}}

\def\deltatran{{\delta}^{\text{tran}}}
\def\tcomp{{l}^{\text{comp}}}

\def\deltacomp{{\delta}^{\text{comp}}}
\def\tedge{{l}^{\text{edge}}}

\def\tedgew{\widehat{l}^{\text{edge}}}

\def\edgearriv{k^{\text{edge}}}

\def\ddl{\tau}

\def\dropedge{e^{\text{edge}}}

\begin{document}
	
	\title{Deep Reinforcement Learning for Task Offloading in Mobile Edge Computing Systems}
	
	\author{Ming Tang
		and~Vincent~W.S.~Wong
		\thanks{Ming Tang and Vincent W.S. Wong are with the Department
			of Electrical and Computer Engineering, The University of British Columbia, Vancouver, Canada.\protect\\
			E-mail: \{mingt,~vincentw\}@ece.ubc.ca}}
	
	\IEEEtitleabstractindextext{
		\begin{abstract}

In mobile edge computing systems, an edge node may have a high load when a large number of mobile devices offload their tasks to it. Those offloaded tasks may experience large processing delay or even be dropped when their deadlines expire. Due to the uncertain load dynamics at the edge nodes,  it is challenging for each device to determine its offloading decision (i.e., whether to offload or not, and which edge node it should offload its task to) in a decentralized manner. In this work, we consider non-divisible and delay-sensitive tasks as well as edge load dynamics, and formulate a task offloading problem to minimize the expected long-term cost. We propose a model-free deep reinforcement learning-based distributed algorithm, where each  device can determine its offloading decision without knowing the task models and offloading decision of other devices. To improve the estimation of the long-term cost in the algorithm, we  incorporate the long short-term memory (LSTM), dueling deep Q-network (DQN), and double-DQN techniques. Simulation results with 50  mobile devices and  five edge nodes show that the proposed algorithm can reduce the ratio of dropped tasks and average task delay by  $86.4\%-95.4\%$ and $18.0\%-30.1\%$, respectively, when compared with several existing algorithms.
	
\end{abstract}
}

\maketitle

\IEEEdisplaynontitleabstractindextext

\IEEEpeerreviewmaketitle

\section{Introduction}
\subsection{Background and Motivation}
Nowadays, mobile devices are responsible for processing more and more computational  intensive tasks, such as data processing, artificial intelligence, and virtual reality. Despite the development of mobile devices, these devices may not be able to process all their tasks locally with a low latency due to their limited computational resources. To facilitate efficient task processing, mobile edge computing (MEC) \cite{mao2017survey}, also known as fog computing \cite{bonomi2012fog}  and multi-access edge computing \cite{porambage2018survey}, is introduced. MEC facilitates mobile devices to offload their computational intensive tasks to nearby edge nodes for processing in order to reduce the task processing delay. It can also reduce the ratio of dropped tasks for those delay-sensitive tasks. 

In MEC, there are two main questions related to  task offloading. The first question is whether a mobile device should offload its task to an edge node or not. The second question is that if a mobile device decides to perform offloading, then which edge node should the device offload its task to. To address these questions, some existing works have proposed task offloading algorithms. Wang \emph{et al.} in \cite{wang2017computation} proposed an algorithm to determine the offloading decisions of the mobile devices to maximize the network revenue. Bi \emph{et al.} in \cite{bi2018computation} focused on a wireless-powered MEC scenario and proposed an algorithm to jointly optimize the offloading and power transfer decisions.  In these works \cite{wang2017computation,bi2018computation}, the processing capacity that each mobile device obtained from an edge node is independent of the number of tasks offloaded to the edge node.

In practice, however, edge nodes may have limited processing capacities, so the processing capacity that an edge node allocated to a mobile device depends on the \emph{load level} at the edge node (i.e.,  number of concurrent tasks offloaded to the edge node). When a large number of mobile devices offload their tasks to the same edge node,  the load at that edge node can be high, and hence those offloaded tasks may experience large processing delay. Some of the tasks may even be dropped when their deadlines expire. Some existing works have addressed the load levels at the edge nodes and proposed centralized task offloading algorithms. Eshraghi \emph{et al.} in \cite{eshraghi2019joint} considered the uncertain computational requirements of the mobile devices, and proposed an algorithm that optimizes the offloading decisions of the mobile devices and the computational resource allocation decision of the edge node. Lyu \emph{et al.} in \cite{lyu2018energy} focused on delay-sensitive tasks and proposed an algorithm to minimize the task offloading energy consumption subject to the task deadline constraint. In \cite{chen2018task},  Chen \emph{et al.} considered a software-defined ultra-dense network, and  designed a centralized algorithm to minimize the task processing delay. In \cite{poularakis2019joint}, Poularakis \emph{et al.} studied the joint optimization of task offloading and routing, taking into account the asymmetric requirements of the tasks. These centralized algorithms in \cite{eshraghi2019joint,lyu2018energy,chen2018task,poularakis2019joint}, however, may require global information of the system (e.g., the arrivals and the sizes of the tasks of all mobile devices) and may incur high signaling overhead.

Other works have proposed distributed task offloading algorithms considering the load levels at the edge nodes, where each mobile device makes its offloading decision in a decentralized manner.  Note that designing such a distributed algorithm is challenging. This is because when a device makes an offloading decision, the device does not know \emph{a priori} the load levels at the edge nodes, since the load also depends on the offloading decisions and task models (e.g., the size and  arrival time of the task) of other mobile devices. In addition, the load levels at the edge nodes may change over time.  To address these challenges, Lyu \emph{et al.} in  \cite{lyu2018distributed} focused on divisible tasks and proposed a Lyapunov-based algorithm to ensure the stability of the task queues. In \cite{li2019incentive}, Li \emph{et al.} considered the strategic offloading interaction among mobile devices and proposed a price-based distributed algorithm. Shah-Mansouri \emph{et al.} in \cite{shah2018hierarchical} designed a potential game-based offloading algorithm to maximize the quality-of-experience of each device. Jo{\v{s}}ilo \emph{et al.} in \cite{jovsilo2019wireless} designed a distributed algorithm based on a Stackelberg game. Yang \emph{et al.} in \cite{yang2018distributed} proposed a distributed offloading algorithm to address the  wireless channel competition among mobile devices. Neto \emph{et al.} in \cite{neto2018uloof} proposed an estimation-based method, where each device makes its offloading decision based on the estimated processing and transmission capacities.

In this work, we focus on the task offloading problem in an MEC system and propose a distributed algorithm that addresses the unknown load level dynamics at the edge nodes. Comparing with the aforementioned works \cite{lyu2018distributed, li2019incentive, shah2018hierarchical, jovsilo2019wireless, yang2018distributed,neto2018uloof}, we consider a different and more realistic MEC scenario. First,  the existing work \cite{lyu2018distributed} considered divisible tasks (i.e., tasks can be arbitrarily divided), which may not be realistic due to the dependency among the bits in a task. On the other hand, although the  works \cite{li2019incentive, shah2018hierarchical, jovsilo2019wireless,yang2018distributed,neto2018uloof} considered non-divisible tasks, they do not take into account the underlying queuing systems. As a result, the processing and transmission of each task should always be accomplished within one time slot (or before the arrival of the next task), which may not  always be guaranteed in practice. Different from those works \cite{lyu2018distributed, li2019incentive, shah2018hierarchical, jovsilo2019wireless,yang2018distributed,neto2018uloof},  we consider non-divisible tasks together with queuing systems and take into account the practical scenario where the processing and transmission of each task can continue for multiple time slots. 
This scenario is challenging to deal with, because for the task of a mobile device arrived in a time slot, its delay can be affected by the decisions of the tasks of other devices arrived in the previous time slots. Second, different from the related works \cite{lyu2018distributed, li2019incentive, shah2018hierarchical, jovsilo2019wireless,yang2018distributed,neto2018uloof} which considered delay-tolerant tasks, we take into account delay-sensitive tasks with processing deadlines. This is challenging to address, because the processing deadlines will affect the load level dynamics at the edge nodes and hence affect the delay of the offloaded tasks.

The model-free deep reinforcement learning (DRL) techniques (e.g., deep Q-learning \cite{mnih2015human}) are candidate methods for solving the task offloading problem in the MEC system, as these methods enable agents to make decisions based on local observations without estimating the dynamics involved in the model. Some existing works such as \cite{huang2019deep,luo2019adaptive,liu2019deep} have proposed DRL-based algorithms for the MEC system, while they focused on centralized offloading algorithms. Zhao \emph{et al.} in \cite{zhao2019deep} proposed a  DRL-based distributed offloading algorithm that addresses the wireless channel competition among mobile devices, while the algorithm at each mobile device requires the quality-of-service information of other mobile devices. Different from those works \cite{huang2019deep,luo2019adaptive,liu2019deep,zhao2019deep}, we aim to propose a  DRL-based  distributed algorithm that can address the unknown load dynamics at the edge nodes. It should also enable each mobile device to make its offloading decision without knowing the  information (e.g., task models, offloading decisions) of other mobile devices.

\subsection{Solution Approach and Contributions}

In this work, we take into account the unknown load level dynamics at the edge nodes and propose a DRL-based distributed offloading algorithm for the MEC system. In the proposed algorithm, each mobile device can determine the offloading decision in a decentralized manner using the information observed locally, including the size of its task, the information of its queues, and the historical load levels at the edge nodes. In addition, the proposed algorithm can handle the time-varying system environments,  including the arrival of new tasks, the computational requirement of each task, and the offloading decisions of other mobile devices. 

The main contributions are as follows. 

\begin{itemize}
	\item \emph{Task Offloading Problem for the MEC System:} 
	We formulate a task offloading problem taking into account the load level dynamics at the edge nodes to minimize the expected long-term cost (considering the delay of the tasks and the penalties for those tasks being dropped). In this problem, we consider non-divisible and delay-sensitive tasks and use queuing systems to model the processing and transmission processes of the tasks. 
	\item \emph{DRL-based Task Offloading Algorithm:} To  achieve the  expected long-term cost minimization considering the unknown load  dynamics at the edge nodes, we propose a model-free DRL-based distributed offloading algorithm that enables each mobile device to make its offloading decision without knowing the task models and offloading decisions of other mobile devices. To improve the estimation of the expected long-term cost  in the proposed algorithm, we incorporate the long short-term memory (LSTM), dueling deep Q-network (DQN), and double-DQN techniques. 
	\item \emph{Performance Evaluation:} We perform simulations and show that when compared with the potential game based offloading algorithm (PGOA) in \cite{yang2018distributed} and the user-level online offloading framework (ULOOF)  in \cite{neto2018uloof}, our proposed DRL-based algorithm can better exploit the  processing capacities of the mobile devices and edge nodes,  and it can significantly reduce the ratio of dropped tasks and the average delay. Under a scenario with 50 mobile devices and  five edge nodes, our proposed algorithm can reduce the ratio of dropped tasks by  $86.4\%-95.4\%$ and reduce the average delay by $18.0\%-30.1\%$ when compared with the existing algorithms.
\end{itemize}

The rest of this paper is organized as follows. The system model is presented in Section \ref{sec:model}, and the problem formulation is given in Section \ref{sec:problem}. We present the DRL-based algorithm in Section \ref{sec:algorithm} and evaluate its performance in Section \ref{sec:exp}. Conclusions are given in Section \ref{sec:conclusion}. For notation, we use $\mathbb{Z}_{++}$ to denote the set of positive integers.

\section{System Model}\label{sec:model}

We consider  a set of edge nodes $\N=\{1,2,\ldots,N\}$ and a set of  mobile devices  $\M=\{1,2,\ldots,M\}$ in an MEC system. The mobile devices can offload their computational tasks to the edge nodes for processing.  We consider one episode that contains a set of time slots $\T=\{1,\ldots,T\}$, where each time slot has a duration of $\Delta$  seconds. In the following, we present the mobile device model and the edge node model, respectively, with an illustration given in Fig. \ref{fig:model1}.

\subsection{Mobile Device Model} 
\begin{figure}[t]
	\centering
	\includegraphics[height=2.9cm]{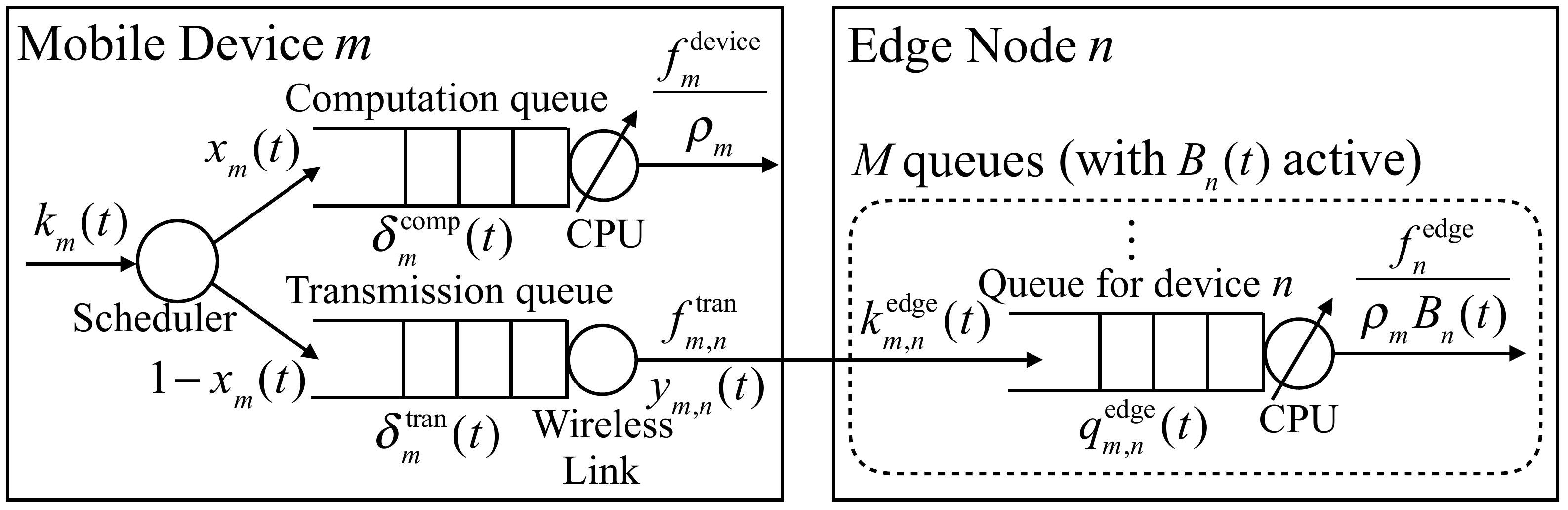}
	\caption{An illustration of an  MEC system with a mobile device $m\in\M$ and an edge node $n\in\N$.}\label{fig:model1}
\end{figure}

During each time slot, we assume that a mobile device either has a new task arrival for processing or does not have a new task arrival. This assumption is reasonable by setting the duration of each time slot to be small, e.g., $\Delta = 0.1$ second. Each mobile device has a scheduler. The scheduler places the newly arrived task to either a computation queue or a transmission queue  (see Fig. \ref{fig:model1}) at the beginning of the next time slot. If the task is placed in the computation queue, then it will be processed locally. If the task is placed in the transmission queue, then it will be sent to an edge node through a wireless link for processing. Note that for the computation (or the transmission) queue, we assume that if the processing (or transmission) of a task is completed in a time slot, then the next task in the queue will be processed (or transmitted) at the beginning of the next time slot. This assumption is consistent with some existing works considering  queuing dynamics in an MEC system (e.g., \cite{liu2016delay-optimal}), and the incurred additional delay can be ignored if the number of time slots that a task needs to wait in the queue and to be processed (or sent) is relatively large. 

In the following, we first present the task model and the task offloading decision, respectively. Then,  we introduce the computation and  transmission queues.

\subsubsection{Task Model} 
At the beginning of time slot  $t\in\T$, if mobile device $m\in\M$ has a newly arrived task to be placed to a queue, then we define a variable  $\task_m(t)\in\mathbb{Z}_{++}$ to denote the unique index of the task. If mobile device $m$ does not have a new task arrival to be placed at the beginning of time slot $t$, then $\task_m(t)$ is set to zero for presentation simplicity.

Let $\arriv_{m}(t)$ (in bits) denote the number of newly arrived bits to be placed in a queue at the beginning of time slot $t\in\T$. If there exists a new task $\task_m(t)$ at the beginning of time slot $t$, then $\arriv_{m}(t)$ is equal to the size of task $\task_m(t)$. Otherwise, $\arriv_{m}(t)$ is set to zero. We set the size of task $\task_m(t)$ to be from a discrete set  $\Lambda\triangleq\{\arriv_{1},\arriv_2,\cdots, \arriv_{{|\Lambda|}}\}$ with $|\Lambda|$ available values. Hence, $\arriv_{m}(t)\in\Lambda\cup\{0\}$. In addition, task $\task_m(t)$ requires a  processing density of $\density_{m}$ (in CPU cycles per bit), i.e., the number of CPU cycles required to process a unit of data.  Task $\task_m(t)$ has a deadline $\ddl_m$ (in time slots). That is, if task $\task_m(t)$ has not been completely processed by the end of time slot $t + \ddl_m - 1$, then it will be dropped immediately.

\subsubsection{Task Offloading Decision}
If mobile device $m\in\M$ has a newly arrived task $\task_{m}(t)$  at the beginning of time slot $t\in\T$, then it needs to make two decisions for task $\task_{m}(t)$. First, the mobile device  decides whether to place the task  to the computation queue or transmission queue. Second, if the task is placed to the transmission queue, then the mobile device decides the edge node to which it should offload the task. 

Let binary variable $x_{m}(t)\in\{0,1\}$ denote whether  task $\task_m(t)$ is scheduled to the computation queue or transmission queue. We set $x_{m}(t)=1$ (or $0$) if the task is  scheduled to the computation queue (or the transmission queue).  At the beginning of time slot $t$, $\arriv_{m}(t) x_{m}(t)$ is the number of bits arrived at the computation queue of mobile device $m$, and $\arriv_{m}(t) (1-x_{m}(t))$ is  the number of bits arrived at the transmission queue of mobile device $m$.

Let binary variable $y_{m,n}(t)\in\{0,1\}$ denote whether task $\task_m(t)$ is offloaded to edge node $n\in\N$ or not. We set $y_{m,n}(t)=1$ if task  $\task_m(t)$  is offloaded to edge node $n$, and $y_{m,n}(t)=0$ otherwise. For presentation convenience, we introduce vector $\boldsymbol{y}_{m}(t) = (y_{m,n}(t),n\in\N)$. We assume that each task can be offloaded to only one edge node, i.e.,
\begin{equation}\label{eq:allocation}
	\sum_{n\in\N}y_{m,n}(t)=\mathbbm{1}(x_{m}(t)=0), ~m\in\M, t\in\T,
\end{equation} 
where  the indicator $\mathbbm{1}(z\in\mathcal{Z})=1$ if $z\in\mathcal{Z}$, and is equal to zero otherwise.

\subsubsection{Computation Queue} 
For mobile device $m\in\M$, let $\iotsev_{m}$ (in CPU cycles) denote its total processing capacity within each time slot. As a result, device $m$ can process a maximum of  $\iotsev_{m}/\density_{m}$ bits  for the computation queue within each time slot, i.e., the total processing capacity in each time slot divided by the required processing density of the tasks.

At the beginning of time slot $t\in\T$, if task  $\task_{m}(t)$  is placed in the computation queue, then we define a variable $\tcomp_{m}(t)\in\T$ to denote the time slot when task $\task_{m}(t)$ has either been processed or  dropped. Without loss of generality, if either task  $\task_{m}(t)$ is not placed in the computation queue or $\task_{m}(t)=0$, then we set $\tcomp_{m}(t)=0$.

Let $\deltacomp_{m}(t)$ (in time slots)  denote the number of time slots that task $\task_{m}(t)$ will wait for processing if it is placed in the computation queue. Note that mobile device $m$ will compute the value of $\deltacomp_{m}(t)$ before it decides the queue to place the task. Given $\tcomp_{m}(t')$ for $t'<t$, the value of $\deltacomp_{m}(t)$ is computed as follows. For $m\in\M$ and $t\in\T$, 
\begin{equation}
	\deltacomp_{m}(t) = 
	\left[\max_{t'\in\{0, 1,\ldots, t-1\}} \tcomp_{m}(t') - t+ 1\right]^+, 
\end{equation}
where the operator $[z]^+ = \max\{0,z\}$, and we set $\tcomp_{m}(0)=0$ for presentation simplicity. Specifically,  the term $\max_{t'\in\{0,1,2,\ldots, t-1\}} \tcomp_{m}(t')$ determines the time slot when all the tasks placed in the computation queue before time slot $t$ has either been processed or dropped. Hence, $\deltacomp_{m}(t) $ determines the number of time slots that task $\task_{m}(t)$ should wait for processing. For example, suppose task $\task_{m}(1)$ is placed in the computation queue, and its processing will be completed in time slot $5$, i.e., $\tcomp_{m}(1)=5$. In the meanwhile, suppose $\task_{m}(2)=0$, i.e., $\tcomp_{m}(2)=0$. At the beginning of time slot $3$,  if  task $\task_{m}(3)$ is placed in the computation queue, then its processing will start after time slot $\tcomp_{m}(1)=5$. Hence, it should wait for $\deltacomp_{m}(3) = [\max\{5, 0\} -3+1]^+= 3$ time slots.

If mobile device $m\in\M$ places task $\task_{m}(t)$ in the computation queue at the beginning of time slot $t\in\T$ (i.e., $x_{m}(t)=1$), then task $\task_{m}(t)$ will have either been  processed or  dropped in time slot $\tcomp_{m}(t)$:
\begin{multline}\label{eq:comp2}
	\tcomp_{m}(t)  =
	\min\Bigg\{t + \deltacomp_{m}(t)  +  \left\lceil\frac{\arriv_{m}(t)}{\iotsev_{m}/\density_{m}}\right\rceil - 1, \\
	t + \ddl_{m}-1\Bigg\},
\end{multline}
where $\lceil \cdot \rceil$ is the ceiling function. Specifically,  the processing of task  $\task_{m}(t)$ will start at the beginning of time slot $t + \deltacomp_{m}(t) $. The number of time slots required to process the task is $ \left\lceil{\arriv_{m}(t)}/({\iotsev_{m}/\density_{m}})\right\rceil$. Hence, the first term in the $\min$ operator is the time slot when the processing of task $\task_{m}(t)$ will be completed without considering the deadline of the task. The second term is the time slot when task $\task_{m}(t)$ will be dropped. As a result, $\tcomp_{m}(t)$ determines the time slot when task $\task_{m}(t)$ will either be processed or dropped.

\subsubsection{Transmission Queue} 
For mobile device $m\in\M$, let $\trans_{m,n}$ (in bits) denote the total transmission capacity from the mobile device to edge node $n\in\N$ in each time slot.\footnote{Since we focus on characterizing the load level dynamics at the edge nodes, we consider a constant transmission capacity in the system model, as in some of the existing works such as \cite{ouyang2018follow,yang2018distributed,jovsilo2019selfish,shah2018hierarchical}.}  At the beginning of time slot $t\in\T$, if task $\task_{m}(t)$ is placed in the transmission queue, then we define a variable $\ttran_{m}(t)\in\T$ to denote the time slot when task $\task_{m}(t)$ has been either sent or dropped. Without loss of generality, if either task $\task_{m}(t)$ is not placed in the transmission queue or  $\task_{m}(t)=0$, then we set $\ttran_{m}(t)=0$. 

Let $\deltatran_{m}(t)$ (in time slots) denote the number of time slots that task $\task_{m}(t)$ should wait for transmission if it is placed in the transmission queue. Note that mobile device $m$ will compute the value of $\deltatran_{m}(t)$ before it has decided on which queue to place the task. Given  $\ttran_{m}(t')$ for $t'<t$,  the value of $\deltatran_{m}(t)$ is computed as follows. For $m\in\M$ and $t\in\T$, 
\begin{equation}
	\deltatran_{m}(t) = 
	\left[\max_{t'\in\{0,1,\ldots, t-1\}} \ttran_{m}(t') - t+ 1\right]^+, 
\end{equation}
where we set $\ttran_{m}(0)=0 $ for presentation simplicity.

If mobile device $m\in\M$ places task $\task_{m}(t)$ in the transmission queue at the beginning of time slot $t\in\T$ (i.e., $x_{m}(t)=0$), then task $\task_{m}(t)$  will either be sent or dropped in time slot $\ttran_{m}(t)$:
\begin{multline}\label{eq:tran2}
	\ttran_{m}(t)  =
	\min\Bigg\{t + \deltatran_{m}(t)  +  \left\lceil\sum_{n\in\N}\frac{y_{m,n}(t)\arriv_{m}(t) }{\trans_{m,n}}\right\rceil - 1, \\
	t + \ddl_{m}-1\Bigg\}.
\end{multline}
The idea of computing  $\ttran_{m}(t)$ in  \eqref{eq:tran2} is similar as that of computing  $\tcomp_{m}(t)$ in \eqref{eq:comp2}.

\subsection{Edge Node Model} 

Each edge node $n\in\N$  maintains $M$ queues, each queue corresponding to a mobile device in set $\M$. We assume that after an offloaded task is received by an edge node in a time slot, the task will be placed in its corresponding queue at the edge node at the beginning of the next time slot. This assumption is used to ensure that a task is processed by an edge node after the task has been completely received, and the incurred additional delay can be ignored if the number of time slots that each task needs for waiting and being processed is relatively large. 

If a task  of mobile device $m\in\M$ is placed  in its corresponding queue at edge node $n\in\N$ at the beginning of time slot $t\in\T$,  then we define a variable  $\edgearriv_{m,n}(t)\in\mathbb{Z}_{++}$ to denote the unique index of the task.\footnote{In each time slot, an edge node receives at most one task from a mobile device, as we have assumed that after a task is sent, the transmission of the next task starts at the beginning of the next time slot.} Specifically, if  task $ \task_m(t')$ for $t'\in\{1,2,\ldots, t-1\}$ is sent to edge node $n$ in time slot $t - 1$,  then we have $\edgearriv_{m,n}(t) = \task_m(t')$. Note that if there does not exist a task of mobile device $m$ being placed in the queue at edge node $n$ at the beginning of time slot $t$, then  we set  $\edgearriv_{m,n}(t) =0$.
Let $\arrivedge_{m,n}(t)\in\Lambda\cup\{0\}$ (in bits) denote the number of bits arrived in the queue of mobile device $m$ at edge node $n$ at the beginning of time slot $t$. If task $\edgearriv_{m,n}(t)$ is placed in the corresponding queue at the beginning of time slot $t$, then $\arrivedge_{m,n}(t)$ is equal to the size of  task $\edgearriv_{m,n}(t)$. Otherwise,  $\arrivedge_{m,n}(t)=0$.

Due to the unknown future load level dynamics at the edge nodes, mobile devices and edge nodes are unaware of the waiting and processing time of the tasks offloaded to the edge nodes until those tasks have  either been processed or dropped. In the following, we first introduce the operation of the queues at the edge nodes. Then, we compute the time slot when a task has either been  processed or dropped.

\subsubsection{Queues at Edge Nodes}
Let  $\Bfog_{m,n}(t)$ (in bits) denote the queue length of mobile device $m\in\M$ at edge node $n\in\N$  at the end of time slot $t\in\T$.  Among those queues at edge node $n$, we refer to the queue of mobile device $m$ as an \emph{active queue} in time slot $t$ if either there is a task of mobile device $m$ arrived at the queue in time slot $t$ (i.e., $\arrivedge_{m,n}(t)>0$) or the queue is non-empty at the end of time slot $t-1$ (i.e., $\Bfog_{m,n}(t-1) > 0$). Let $\actset_{n}(t)$ denote the set of active queues at edge node $n$ in time slot $t$. That is, for $n\in\N$ and $t\in\T$,
\begin{equation}\label{eq:fog-share}
	\actset_{n}(t) = \{m~|~m\in\M, \arrivedge_{m,n}(t)> 0~\text{or}~\Bfog_{m,n}(t-1) >0\}.
\end{equation} 
Let $\actnum_{n}(t)$ denote the number of active queues at edge node $n$ in time slot $t$, i.e., $\actnum_{n}(t)=|\actset_{n}(t)|$.

Within time slot $t\in\T$,  the active  queues at an edge node $n\in\N$,  i.e., the  queues in set $\actset_{n}(t)$, equally share the processing capacity of edge node $n$. This is a generalized processor sharing (GPS) model \cite{parekh1993generalized} with equal processing capacity sharing, and it can be approximated by practical algorithms such as the fair queuing algorithm in \cite{demers1989analysis}. Let $\fogsev_{n}$ denote the total processing capacity of edge node $n$ within each time slot. Therefore, edge node $n$ can process a maximum of $\fogsev_{n}/(\density_{m}\actnum_{n}(t))$ bits for any active device $m\in\actset_{n}(t)$ within time slot $t$. 

For any queue at the edge nodes, we assume that if the processing of a task is completed in a time slot, then the next task in the queue will be processed at the beginning of the next time slot.  To compute the  queue length of mobile device $m\in\M$ at edge node $n\in\N$, let $\dropedge_{m,n}(t)$ (in bits) denote the number of bits of the tasks dropped by the queue at the end of time slot $t\in\T$. Consequently, the queue length  $\Bfog_{m,n}(t)$ is updated as follows.  For  $m\in\M$, $n\in\N$, and $t\in\T$,
\begin{multline}\label{eq:buf-fog1}
	\Bfog_{m,n}(t) =\Bigg[\Bfog_{m,n}(t-1)  +\arrivedge_{m,n}(t)   \\- \frac{\fogsev_{n}}{\density_{m} \actnum_{n}(t)} \mathbbm{1}\left(m\in\actset_{n}(t)\right)- \dropedge_{m,n}(t)\Bigg]^+.
\end{multline}
Intuitively, the queue length  $\Bfog_{m,n}(t)$ is equal to the queue length in the previous time slot $\Bfog_{m,n}(t-1)$ plus the difference between the bits arrived in time slot $t$ and the bits being served (i.e., processed or dropped) in time slot $t$.

\subsubsection{Task Processing or Dropping}
If task $\edgearriv_{m,n}(t)$ of mobile device $m\in\M$ is placed in the corresponding queue at edge node $n\in\N$ at the beginning of time slot $t\in\T$, then we define a variable $\tedge_{m,n}(t)\in\T$ to denote the time slot when this task  has either been  processed or dropped by edge node $n$. Due to the uncertain future load at edge node $n$, the value of $\tedge_{m,n}(t)$ is unknown to mobile device $m$ and edge node $n$ until the associated task $\edgearriv_{m,n}(t)$ has  either been processed or dropped. Without loss of generality, if  $\edgearriv_{m,n}(t)=0$, then we set $\tedge_{m,n}(t)=0$.

For the definition of variable $\tedge_{m,n}(t)$, let $\tedgew_{m,n}(t)$ denote the time slot when the processing of task $\edgearriv_{m,n}(t)$  starts, i.e., for $m\in\M$, $n\in\N$, and $t\in\T$,
\begin{equation}
	\tedgew_{m,n}(t) = \max\left\{t,\max_{t'\in\{0,1,\ldots, t-1\}}\tedge_{m,n}(t') + 1\right\},
\end{equation}
where we set $\tedge_{m,n}(0)=0$. Specifically, the time slot  when the processing of task $\edgearriv_{m,n}(t)$  starts should be no earlier than the time slot that the task is placed in the queue or the time slot when each of the tasks arrived earlier has been processed or dropped.

Given  the realization of the load levels at edge node $n$,  $\tedge_{m,n}(t)$ is  the time slot satisfying the following constraints. For $m\in\M$, $n\in\N$, and $t\in\T$,
\begin{equation}\label{eq:edge3}
	\sum_{t' = \tedgew_{m,n}(t)}^{\tedge_{m,n}(t)} \frac{\fogsev_{n}}{\density_{m}\actnum_{n}(t')} \mathbbm{1}\left(m\in\actset_{n}(t')\right) \geq  \arrivedge_{m,n}(t), 
\end{equation}
\begin{equation}\label{eq:edge4}
	\sum_{t' = \tedgew_{m,n}(t)}^{\tedge_{m,n}(t) - 1} \frac{\fogsev_{n}}{\density_{m}\actnum_{n}(t')} \mathbbm{1}\left(m\in\actset_{n}(t')\right)< \arrivedge_{m,n}(t).
\end{equation}
Specifically,  the total processing capacity that edge node $n$ allocated to mobile device $m$ from time slot $\tedgew_{m,n}(t)$ to time slot $\tedge_{m,n}(t)$  should be no smaller than the size of task $\edgearriv_{m,n}(t)$, while the corresponding total processing capacity allocated from time slot $\tedgew_{m,n}(t)$ to time slot $\tedge_{m,n}(t) -1 $ should be  smaller than the size of the task.

\section{Task Offloading Problem in MEC}\label{sec:problem}
In this section, we present the task offloading problem for the MEC system. Specifically, at the beginning of each time slot, each mobile device observes its state (e.g., task size, queue information). If there is a newly arrived task to be processed, then the mobile device chooses an action for the task. The observed state and the chosen action will result in a cost (i.e., the delay of the task if the task is processed, or a penalty if it is dropped) for the mobile device. The objective of each mobile device is to minimize its expected long-term cost by optimizing the policy mapping from states to actions. In the following, we first introduce the state, action, and cost, respectively. We then formulate the cost minimization problem for each device.

\subsection{State}\label{subsec:state}
At the beginning of time slot $t\in\T$, each device $m\in\M$ observes its state information, including the task size, the information related to the queues, and the load level history at the edge nodes. Specifically, mobile device $m$ maintains the following state vector: 
\begin{equation}\label{eq:state}
	\s_{m}(t) = \Big(\arriv_{m}(t), \deltacomp_{m}(t),  \deltatran_{m}(t),  
	\boldBfog_{m}(t-1),\hiscon(t)\Big),\vspace{-2mm}
\end{equation}
where vector $\boldBfog_{m}(t-1)=(\Bfog_{m,n}(t-1),n\in\N)$. The matrix $\hiscon(t)$ includes the history of the load level (i.e., the number of active queues) of each edge node within the previous $\ts$ time slots (i.e., from time slot $t-\ts$ to time slot $t-1$), with which the load levels at the edge nodes in the near future can be estimated. It is a matrix with size $\ts\times N$. Let  $\{\hiscon(t)\}_{i,j}$ denote the $(i,j)$ element of matrix $\hiscon(t)$, and it corresponds to the load level history of edge node $j$ in the $i^{\text{th}}$ time slot starting from time slot $t-\ts$, i.e., time slot $t-\ts+i-1$. The element $\{\hiscon(t)\}_{i,j} $ is defined as follows:
\begin{equation}
	\left\{\hiscon(t)\right\}_{i,j} =  \actnum_{j}(t-\ts+i-1),
\end{equation}
which is the number of active queues of edge node $j$ in  time slot $t-\ts+i-1$. Let $\mathcal{S}$ denote the discrete and finite state space of each mobile device.  Formally, set $\mathcal{S} = \Lambda\times \{0,1,\ldots,T\}^2\times\mathcal{Q}^{N}\times \{0,1,\ldots, M\}^{\ts\times N}$, where $\mathcal{Q}$ denotes the set of the  available  values of the queue length at an edge node within the $T$ time slots.

Mobile device $m\in\M$ can obtain state information $\arriv_{m}(t)$,  $ \deltacomp_{m}(t)$, and $\deltatran_{m}(t)$ through local observation at the beginning of time slot $t$. For state information $\boldBfog_{m}(t-1)$, mobile device $m$ can compute this vector according to the number of bits of device $m$ transmitted to each edge node in each time slot and the number of bits of device $m$ processed or being dropped by each edge node  in each time slot according to \eqref{eq:buf-fog1}.  For matrix $\hiscon(t)$, we assume that each edge node will broadcast its number of active queues at the end of each time slot. Since the number of active queues is always a small number, which can be represented by several bits, the broadcasting will only incur a small signaling overhead.

\subsection{Action}
At the beginning of time slot $t\in\mathcal{T}$, if mobile device $m\in\mathcal{N}$ has a new task arrival $\task_{m}(t)$, then it will choose the actions for task $\task_{m}(t)$: (a) whether to schedule the task  to the computation queue or  the transmission queue, i.e., $x_{m}(t)$; (b) which edge node the task is offloaded to, i.e., $\boldsymbol{y}_{m}(t) = (y_{m,n}(t),n\in\N)$.  Hence, the action of device $m$ in time slot $t$ is represented by the following action vector:
\begin{equation}
	\a_{m}(t) = \left(x_{m}(t), \boldsymbol{y}_{m}(t)\right).
\end{equation}
Let $\As$ denote the decision space of each mobile device, i.e., $\As = \{0,1\}^{1+N}$.

\subsection{Cost} 
Let $\delay_{m}(\s_{m}(t),\a_{m}(t))$ (in time slots) denote the delay of task $\task_{m}(t)$, given the observed state $\s_{m}(t)$ and the selected action $\a_{m}(t)$. For $m\in\M$ and $t\in\T$, if $x_m(t)=1$, then
\begin{equation}\label{eq:delay1}
	\delay_{m}(\s_{m}(t),\a_{m}(t)) = 	\tcomp_{m}(t) - t + 1;
\end{equation}
if $x_m(t)=0$, then
\begin{multline}\label{eq:delay2}
	\delay_{m}(\s_{m}(t),\a_{m}(t)) \\=\sum_{n\in\N}\sum_{t'=t}^{T}\mathbbm{1}(\edgearriv_{m,n}(t') = \task_m(t))\tedge_{m,n}(t')  - t + 1.
\end{multline}
Specifically, the delay of task $\task_{m}(t)$ is the number of time slots between time slot $t$ and the time slot when task $\task_{m}(t)$ has either been  processed or dropped. 

There is a cost $c_{m}(\s_{m}(t),\a_{m}(t))$ associated with  task $\task_{m}(t)$. If task $\task_{m}(t)$ has been processed, then
\begin{equation}\label{eq:cost1}
	c_{m}(\s_{m}(t),\a_{m}(t)) =  \delay_{m}(\s_{m}(t),\a_{m}(t)).
\end{equation}
On the other hand, if task $\task_{m}(t)$ has been dropped, then
\begin{equation}\label{eq:cost2}
	c_{m}(\s_{m}(t),\a_{m}(t)) = C,
\end{equation}
where $C> 0$ is a  constant penalty. Without loss of generality, if task $\task_{m}(t)=0$, then we set $c_{m}(\s_{m}(t),\a_{m}(t))=0$. In the remaining part of this work, we use the short form $c_{m}(t)$ to denote $c_{m}(\s_{m}(t),\a_{m}(t))$.

\subsection{Problem Formulation}
A policy of device $m\in\M$ is a mapping from its state to its action, i.e., $\pi_{m}: \mathcal{S} \rightarrow \mathcal{A}$. We aim to find the optimal policy $\pi_{m}^*$ for each device $m$ such that its expected long-term cost is minimized, i.e.,
\begin{equation}\label{eq:opt-drl}
	\begin{aligned}
		\pi_{m}^* = \arg\mathop{\text{minimize}}_{\pi_{m}} \quad & \mathbb{E}\left[\left.\sum_{t\in\T} \gamma^{t-1}c_{m}(t)~\right|~\pi_{m} \right]\\
		\textrm{subject to} ~~~& \text{constraints }  \eqref{eq:allocation}-\eqref{eq:tran2}, \eqref{eq:buf-fog1}-\eqref{eq:edge4},\\
		\qquad ~& \eqref{eq:delay1}- \eqref{eq:cost2},\\
	\end{aligned}
\end{equation}
where $\gamma\in(0,1]$ is a discount factor that characterizes the discounted cost in the future. The expectation $\mathbb{E}[\cdot]$ is with respect to the time-varying system environments, including the task arrivals and the computational requirements of the tasks of all mobile devices as well as the offloading decisions of the mobile devices other than device $m$.

Solving problem \eqref{eq:opt-drl} is challenging. This is mainly due to the unknown load levels at the edge nodes, which depend on the decisions and the task models (e.g., the size and  arrival time of the task) of other mobile devices, as well as the unknown future task models of the device itself. In this work, we propose a DRL-based offloading algorithm that addresses the challenge by learning the mapping from each state-action pair to its expected long-term cost.

\section{DRL-Based Offloading Algorithm}\label{sec:algorithm}
In this section, we propose a DRL-based offloading algorithm that enables the distributed offloading decision making of each mobile device. In the proposed algorithm, each mobile device  aims to learn a mapping from each state-action pair to a Q-value, which characterizes the expected long-term cost of the state-action pair. The mapping is determined by a neural network. Based on the mapping, each device can select the action inducing the minimum Q-value under its  state to minimize its expected long-term cost.  

In the following, we first introduce the neural network for a mobile device  that characterizes its mapping from state-action pairs to Q-values. Then, we present the DRL-based algorithm and describe the message exchange between a mobile device and an edge node.

\subsection{Neural Network}\label{subsec:neural}
\begin{figure}
	\centering
	\includegraphics[height=4.5cm]{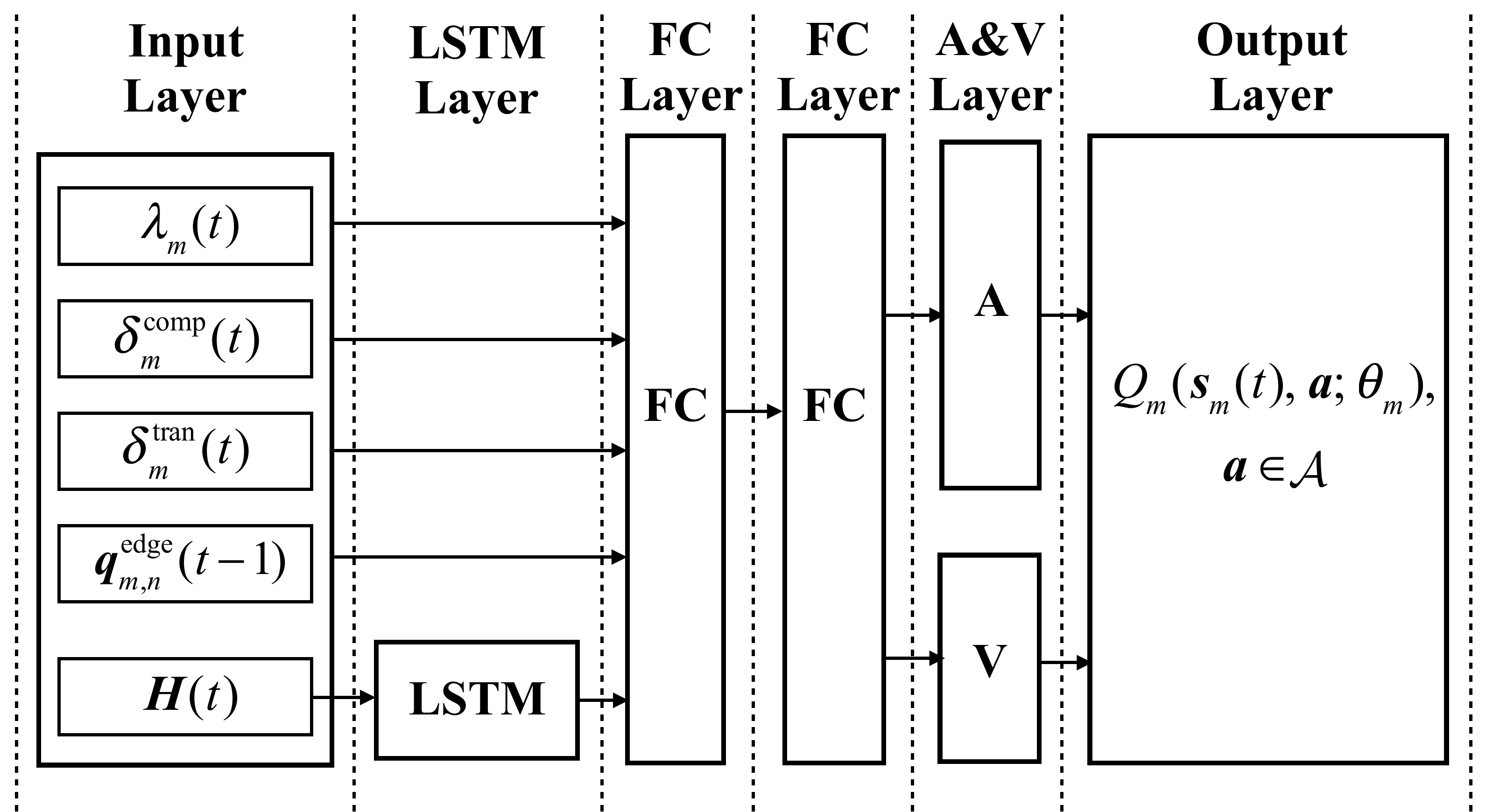}
	\caption{The neural network of mobile device $m\in\M$ with parameter vector $\boldsymbol{\theta}_m$, which maps from state $\s_m(t)\in\Ss$ to the Q-value of each action $\a\in{\As}$.}\label{fig:DNN}
\end{figure}

The objective of the neural network is to find a mapping from each state to a set of Q-values, each corresponding to an action. As shown in Fig. \ref{fig:DNN}, for any mobile device $m\in\M$, we consider a neural network with six layers: an input layer, an LSTM layer, two fully connected (FC) layers, an advantage and value (A\&V) layer, and an output layer. Let  $\boldsymbol{\theta}_m$ denote the parameter vector  of the neural network of device $m$, which includes the weights of all connections and the biases of all neurons from the input layer to the A\&V layer.\footnote{The weights of the connections between the A\&V layer and the output layer as well as  the bias of the neurons in the output layer are  given and fixed. Hence, we do not include them in the network parameter vector $\boldsymbol{\theta}_m$, as the vector $\boldsymbol{\theta}_m$ includes the parameters that are adjustable through learning in the DRL-based algorithm.}  The details of each layer  are as follows. 

\subsubsection{Input Layer} 
This layer is responsible for taking the state vector as input and passing them to the following layers. For mobile device $m\in\M$, the state information includes $\arriv_{m}(t)$, $ \deltacomp_{m}(t)$, $\deltatran_{m}(t)$,  $\boldBfog_{m}(t-1)$, and $\hiscon(t)$. The state information  $\arriv_{m}(t)$, $ \deltacomp_{m}(t)$, $\deltatran_{m}(t)$, and $\boldBfog_{m}(t-1)$ will be passed to the FC layer, and  $\hiscon(t)$ will be passed to the LSTM layer.

\subsubsection{LSTM Layer} This layer is responsible for learning the dynamics of  the load levels at the edge nodes. This is achieved by including an  LSTM network\cite{hochreiter1997long,gers1999learning}. We use the LSTM network because  it can keep track of the state $\hiscon(t)$ over time. It can  provide the neural network the ability of estimating the load levels at the edge nodes in the future using the history. 

\begin{figure}
	\centering
	\includegraphics[height=4.25cm]{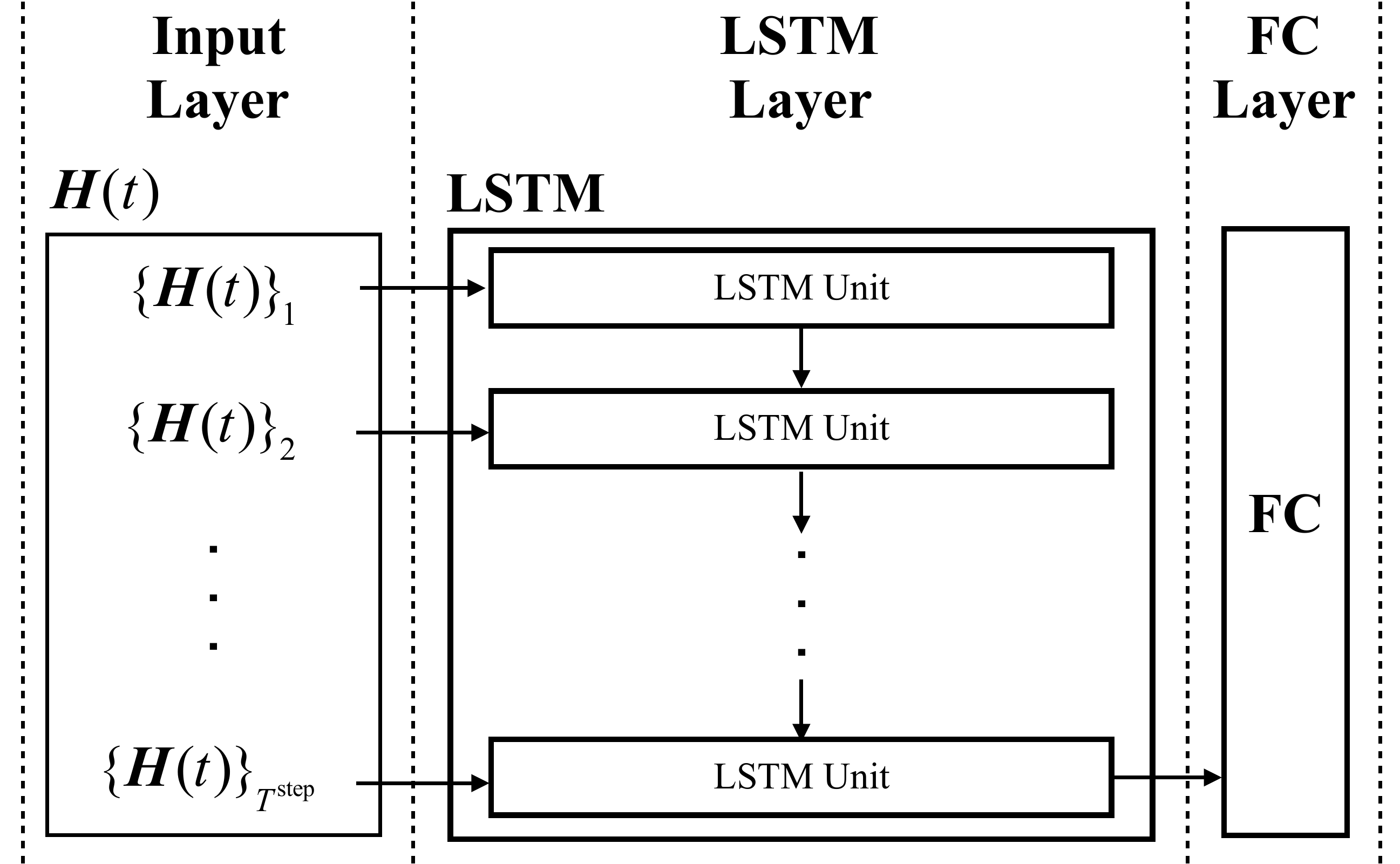}
	\caption{An LSTM network with $\ts$ LSTM units.
	}\label{fig:LSTM}
\end{figure}

Specifically, the LSTM network takes the matrix $\hiscon(t)$ as input so as to learn the load level dynamics.  Fig. \ref{fig:LSTM} shows the structure of an LSTM network. The LSTM network  contains $\ts$ LSTM units, each of which contains a set of hidden neurons. Each LSTM unit takes one row of the matrix $\hiscon(t)$ as input, we let $\{\hiscon(t)\}_{i}$ denote the $i^{\text{th}}$ row of  matrix $\hiscon(t)$ in Fig. \ref{fig:LSTM}.  These LSTM units are connected in sequence so as to keep track of the variations of the sequences from $\{\hiscon(t)\}_{1}$ to $\{\hiscon(t)\}_{\ts}$, which can reveal the variations of the load levels at the edge nodes among time slots. The LSTM network will output the information that indicates the dynamics of the load levels in the future in the last LSTM unit, where the output will be connected to the neurons in the next layer for further learning.

\subsubsection{FC Layers} 
The two FC layers are responsible for learning the mapping from the state and the learned load level dynamics to the Q-values of the actions. Each FC layer contains a set of neurons with rectified linear unit (ReLU). For the first FC layer, the input of each neuron  connects to the neurons in the input layer corresponding to all the states (except the matrix $\hiscon(t)$) and the LSTM network in the LSTM layer. The output of each neuron connects to each of the neurons in the second FC layer. For the second FC layer, the output of   each neuron connects to each of the neurons in  the A\&V layer.

\subsubsection{A\&V Layer and Output Layer} The  A\&V layer and the output layer implement the dueling-DQN technique \cite{wang2015dueling} and determine the Q-value of each action as output. The main idea of the dueling-DQN is to first separately learn a \emph{state-value} (i.e., the portion of the Q-value resulting from the state) and \emph{action-advantage values} (i.e.,  the portion of the Q-value resulting from the actions), and then use the state-value and action-advantage values to determine the Q-values of state-action pairs. This technique can improve the estimation of the Q-values through separately evaluating the expected long-term cost resulting from a state and an action.

The A\&V layer contains two networks, denoted by network A and network V (see Fig. \ref{fig:DNN}). The network A contains an FC network with a set of neurons. It is responsible for learning the action-advantage value  of each action $\a\in\As$. For mobile device $m\in\M$, let $A_m(\s_m(t),\a;\boldsymbol{\theta}_m)$ denote the action-advantage value of action $\a$ under state $\s_m(t)\in\mathcal{S}$ with network parameter vector $\boldsymbol{\theta}_m$. The network V contains an FC network  with a set of neurons. It is responsible for learning the state-value. For mobile device $m$, let $V_m(\s_m(t);\boldsymbol{\theta}_m)$ denote state-value of state $\s_m(t)$ with network parameter vector $\boldsymbol{\theta}_m$.  The values of $A_m(\s_m(t),\a;\boldsymbol{\theta}_m)$  and $V_m(\s_m(t);\boldsymbol{\theta}_m)$ are determined by the  parameter vector $\boldsymbol{\theta}_m$ and the neural network structure from the input layer to the A\&V layer, where vector $\boldsymbol{\theta}_m$ is adjustable and will be trained in the DRL-based algorithm. 

Based on the A\&V layer,   for mobile device $m\in\M$, the resulting Q-value  of action $\a\in\As$ under state $\s_m(t)\in\Ss$ in the output layer is given as follows\cite{wang2015dueling}:
\begin{multline}
	Q_m(\s_m(t),\a;\boldsymbol{\theta}_m) = V_m(\s_m(t);\boldsymbol{\theta}_m) + \Bigg(A_m(\s_m(t),\a;\boldsymbol{\theta}_m) 
	\\ - \frac{1}{|\As|}\sum_{\a'\in\As}A_m(\s_m(t),\a';\boldsymbol{\theta}_m) \Bigg),
\end{multline}
which is the sum of the state-value under the corresponding state and the additional action-advantage value of the corresponding action (i.e., the difference between the action-advantage value of the action and the average action-advantage value over all  actions).

In summary, from the input layer to the output layer, the neural network of mobile device $m\in\M$ with parameter vector $\boldsymbol{\theta}_m$ forms a mapping from state-action pairs to Q-values (i.e., under any observed state $\s_m(t)\in\Ss$, there is a Q-value for each action $\a\in\As$, denoted by $Q_m(\s_m(t),\a;\boldsymbol{\theta}_m)$), which characterizes the expected long-term cost under the observed state and each of the actions in the action space.

\subsection{DRL-Based Algorithm}
In our proposed DRL-based algorithm, we let edge nodes help mobile devices to train the neural network to alleviate the computational loads at the mobile devices. Specifically, for each mobile device $m\in\M$, there is an edge node $n_{m}\in\N$ which helps device $m$ with the training. This edge node $n_{m}$ can be the edge node that has the maximum transmission capacity with mobile device $m$. For presentation convenience, let $\M_{n}\subset\M$ denote the set of mobile devices whose training is performed by edge node $n\in\N$, i.e., $\M_{n}=\{m\in\M~|~{n}_{m} = n\}$. Note that it is reasonable to let edge nodes help with the training directly. This is because the information exchange involved in the training (including the state information and the neural network parameter) is small. In addition, the required processing capacity of the training in each time slot can be much less than those of the tasks of mobile devices. 

The DRL-based algorithm to be executed at mobile device $m\in\M$ and  edge node $n\in\N$ are given in Algorithms \ref{alg:drl1} and \ref{alg:drl2}, respectively. The key idea of the  algorithm is to train the neural network using the experience\footnote{We use the term ``experience" to refer to the tuple consisting of state, action, cost, and next state, as used in \cite{wang2015dueling,mnih2013playing}. Alternatively, some other existing works, such as \cite{lillicrap2015continuous,mnih2015human}, used the term ``transition".} (i.e., state, action, cost, and next state) of the mobile device to obtain the mapping from state-action pairs to Q-values, based on which the device can select the action leading to the minimum Q-value under the observed state to minimize its expected long-term cost.

\begin{figure}[t]
	\vspace{-4.65mm}
	\begin{minipage}{\linewidth}
		\begin{algorithm}[H]
			\caption{DRL-based Algorithm at Device $m\in\M$}\label{alg:drl1}
			\small
			\begin{algorithmic}[1]
				\FOR{episode from 1 to \emph{\#\_of\_Episodes}} 
				\STATE Initialize $\s_{m}(1)$;\
				\FOR{time slot $t\in\T$}
				\IF{device $m$ has a new task arrival $\task_m(t)$}
				\STATE Send a \emph{parameter\_request} to edge node $n_{m}$;\
				\STATE Receive network parameter vector $\thetae_{m}$;\
				\STATE Select an action $\a_{m}(t)$ according to \eqref{eq:action};\
				\ENDIF
				\STATE Observe the next state $\s_{m}(t+1)$;\
				\STATE Observe a set of costs $\{c_{m}(t'),~t'\in\widetilde{\T}_{m,t}\}$;\
				\FOR{each task $\task_m(t')$ with $t'\in\widetilde{\T}_{m,t}$}
				\STATE Send $(\s_{m}(t'),\a_{m}(t'),c_{m}(t'),\s_{m}(t'+1))$ to $n_{m}$;\
				\ENDFOR
				\ENDFOR
				\ENDFOR
			\end{algorithmic}
		\end{algorithm}
	\end{minipage}
\end{figure}

In the DRL-based algorithm, the edge node $n\in\N$ maintains a replay memory $D_{m}$ and two neural networks for device $m\in\M_{n}$. The replay memory $D_{m}$  stores the observed experience $(\s_{m}(t), \a_{m}(t), c_{m}(t), \s_{m}(t+1))$ of mobile device $m$ for some $t\in\T$, where we refer $(\s_{m}(t), \a_{m}(t), c_{m}(t), \s_{m}(t+1))$ as experience $t$ of mobile device $m$. The experience in the replay memory is used to train the neural networks. The two neural networks include an \emph{Eval\_Net$_{m}$}  and a \emph{Target\_Net$_{m}$}, and their Q-values are represented by $\Qe_{m}(\s_m(t),\a;\thetae_{m})$ and $\Qt_{m}(\s_m(t),\a;\thetat_{m})$ under observed state $\s_m(t)\in\Ss$ and action $\a\in\As$, respectively. Note that the \emph{Eval\_Net$_{m}$}   and the \emph{Target\_Net$_{m}$} have the same neural network structure, as presented in Section \ref{subsec:neural}, while they have different network parameter vectors $\thetae_{m}$ and $\thetat_{m}$, respectively. The \emph{Eval\_Net$_{m}$} is used for action selection. The \emph{Target\_Net$_{m}$} is used for characterizing a target Q-value, which approximates the expected long-term cost of an action under the observed state. This target Q-value will be used for updating the network parameter vector $\thetae_{m}$ in \emph{Eval\_Net$_{m}$} by minimizing the difference between the Q-value under \emph{Eval\_Net$_{m}$}  and the target Q-value. The  initialization of the replay memory $D_{m}$ and two neural networks are given  in steps 1$-$3 in Algorithm \ref{alg:drl2}.

\begin{algorithm}[t]
	\caption{DRL-Based Algorithm at Edge Node $n\in\N$}\label{alg:drl2}
	\small
	\begin{algorithmic}[1]
		\STATE Initialize replay memory $D_{m}$ for each device $m\in\M_{n}$ and set $\text{Count} := 0$;\
		\STATE Initialize  \emph{Eval\_Net$_{m}$} with random parameter $\thetae_{m}$ for each device $m\in\M_{n}$;\ 
		\STATE Initialize \emph{Target\_Net$_{m}$} with random parameter $\thetat_{m}$ for each device $m\in\M_{n}$;\ 
		\WHILE{True}
		\IF{receive a \emph{parameter\_request} from  device $m\in\M_{n}$}
		\STATE Send $\thetae_{m}$ to device $m$;\
		\ENDIF
		\IF{receive an experience $(\s_{m}(t),\a_{m}(t),c_{m}(t),\s_{m}(t+1))$ from device $m\in\M_{n}$}
		\STATE Store $(\s_{m}(t),\a_{m}(t),c_{m}(t),\s_{m}(t+1))$ in $D_{m}$;\
		\STATE Sample a set of experiences (denoted by $\mathcal{I}$) from  $D_{m}$;\
		\FOR{each experience $i\in\I$}
		\STATE Obtain experience  $(\s_{m}(i),\a_{m}(i),c_{m}(i),\s_{m}(i+1))$;\
		\STATE Compute $\qtmp_{m,i}$ 
		according to \eqref{eq:q-target1};\
		\ENDFOR
		\STATE Set vector $\Qtmp_{m}:=(\qtmp_{m,i},i\in\I)$;\
		\STATE Update $\thetae_{m}$ to minimize $L(\thetae_{m},\Qtmp_{m})$ in \eqref{eq:loss};\ 
		\STATE $\text{Count} := \text{Count}+ 1$;\
		\IF{$\text{mod(Count, Replace\_Threshold)} = 0$}
		\STATE $\thetat_{m}:= \thetae_{m}$;\
		\ENDIF
		\ENDIF
		\ENDWHILE
	\end{algorithmic}
\end{algorithm}

In the following, we present the DRL-based algorithm at mobile device $m\in\M$ and edge node $n\in\N$, respectively.

\subsubsection{Algorithm \ref{alg:drl1} at Mobile Device $m\in\M$}
We consider multiple  episodes, where \emph{\#\_of\_Episodes} denotes the number of  episodes.  At the beginning of each episode,  mobile device $m\in\M$ initializes the state, i.e., 
\begin{equation}\s_{m}(1)= (\arriv_{m}(1), \deltacomp_{m}(1), \deltatran_{m}(1), \boldBfog_{m}(0),\hiscon(1)),\end{equation}
where we set $\Bfog_{m,n}(0)=0$ for all $n\in\N$, and $\hiscon(1)$ is a zero matrix with size $\ts\times N$.\footnote{For matrix $\hiscon(t)$, for any $t-1<\ts$ (i.e., the number of observed history is smaller than $\ts$), $\{\hiscon(t)\}_{i}=\boldsymbol{0}$ for $i=1,2,\ldots,\ts-(t-1)$, where $\boldsymbol{0}$ is a zero vector with size $N$.} Each episode contains a set of time slots $\T$.

At the beginning of time slot $t\in\T$, if mobile device $m$ has a new task arrival $\task_{m}(t)$, then it will send a \emph{parameter\_request} to edge node $n_{m}$. Upon receiving the requested  parameter vector $\thetae_{m}$ of \emph{Eval\_Net$_{m}$} from edge node  $n_{m}$,  device $m$ will choose its action for task $\task_m(t)$ as follows:
\begin{multline}\label{eq:action}
	\a_{m}(t) =\\
	\left\{\begin{array}{ll}\!\!\!
		\text{select a random action from $\As$},&\!\!\!\! \text{with prob. $\epsilon$}, \\
		\!\!\!\arg \min_{\a\in\As} \Qe_{m}(\s_{m}(t),\a;\thetae_{m}),&\!\!\!\!\text{with   prob.  $1-\epsilon$},
	\end{array}\right.
\end{multline}
where `prob.' is the short-form for probability, and $\epsilon$ is the probability of random exploration. The value of $\Qe_{m}(\s_{m}(t),\a;\thetae_{m})$ is the Q-value under the current parameter $\thetae_{m}$ of neural network \emph{Eval\_Net$_{m}$}. Intuitively, with a probability $1-\epsilon$, the mobile device chooses the action that corresponds to the minimum Q-value under the observed state $\s_{m}(t)$ based on \emph{Eval\_Net$_{m}$}.

At the beginning of the next time slot (i.e., time slot $t+1$), mobile device $m$ observes the next state $\s_{m}(t+1)$. On the other hand, as the processing and the transmission of a task may continue for multiple time slots, the cost $c_{m}(t)$, which depends on the delay of task $\task_m(t)$, may not be observed at the beginning of time slot $t+1$. Instead, mobile device $m$ may observe a set of costs belonging to some tasks $\task_m(t')$ with time slot $t'\leq t$. To address this, for device $m$, we define  $\widetilde{\T}_{m,t}\subset\T$ as the set of time slots such that each task $\task_m(t')$ associated with time slot $t'\in\widetilde{\T}_{m,t}$ has been processed or dropped in time slot $t$. Set $\widetilde{\T}_{m,t}$ is defined as follows:
\begin{multline}\label{eq:rewardtime}
	\widetilde{\T}_{m,t} = \Bigg\{t'~\Bigg|~t'=1,2,\ldots, t,~ \arriv_{m}(t')>0,~ x_m(t')\tcomp_{m}(t')\\+
	(1 -x_m(t')) \sum_{n\in\N}\sum_{i=t'}^{t}\mathbbm{1}(\edgearriv_{m,n}(i) = \task_m(t'))\tedge_{m,n}(i) = t \Bigg\}.
\end{multline} 
In \eqref{eq:rewardtime}, $\arriv_{m}(t')>0$ implies that there is a newly arrived task $\task_m(t')$ in time slot $t'$. Specifically, set $\widetilde{\T}_{m,t}$ contains a time slot $t'\in\{1,2,\ldots, t\}$ if task $\task_m(t')$ has been processed or dropped in time slot $t$. Hence, at the beginning of time slot $t+1$, mobile device $m$ can observe a set of costs $\{c_{m}(t'), t'\in\widetilde{\T}_{m,t}\}$, where set $\widetilde{\T}_{m,t}$ can be an empty set for some $m\in\M$ and $t\in\T$. Then, for each task $\task_m(t')$ with $t'\in\widetilde{\T}_{m,t}$,   device $m$ sends its experience $(\s_{m}(t'), \a_{m}(t'), c_{m}(t'), \s_{m}(t'+1))$ to edge node $n_{m}$.

\subsubsection{Algorithm \ref{alg:drl2} at Edge Node $n\in\N$}
After initializing the replay memory $D_{m}$ as well as the neural networks \emph{Eval\_Net$_{m}$} and  \emph{Target\_Net$_{m}$} for device $m\in\M_{n}$, edge node $n\in\N$ will wait for the request messages from the mobile devices in set $\M_{n}$. If edge node $n$ receives a \emph{parameter\_request} from mobile device $m\in\M_{n}$, then it will send the current parameter vector $\thetae_{m}$ of \emph{Eval\_Net$_{m}$} to device $m$. On the other hand, if edge node $n$  receives  an experience $(\s_{m}(t), \a_{m}(t), c_{m}(t), \s_{m}(t+1))$ from mobile device $m\in\M_{n}$,  then it will store the experience  in memory $D_{m}$. The memory has a maximum size, and it serves in a first-in first-out (FIFO) manner. Note that we do not require synchronization between mobile device $m$ and its associated edge node $n_{m}$. The edge node will train the neural network (in steps 10$-$20 in Algorithm \ref{alg:drl2}) to update the  parameter vector $\thetae_{m}$ of \emph{Eval\_Net$_{m}$}  as follows.   

The edge node will randomly sample a set of experiences from the memory (in step 10), denoted by $\mathcal{I}$.
Based on these experience samples, the key idea of the update of \emph{Eval\_Net$_{m}$}  is to minimize the difference between the Q-values  under  \emph{Eval\_Net$_{m}$}  and the target Q-values computed based on the experience samples under \emph{Target\_Net$_{m}$}. Specifically, for the experience samples in set $\I$, the edge node will compute $\Qtmp_{m}=(\qtmp_{m,i},i\in\I)$ and update $\thetae_{m}$ in \emph{Eval\_Net$_{m}$} by minimizing the following loss function: 
\begin{multline}\label{eq:loss}
	L(\thetae_{m},\Qtmp_{m})  = \frac{1}{|\I|}\sum_{i\in\I}\Big(\Qe_{m}(\s_{m}(i),\a_{m}(i);\thetae_{m})
	\\- \qtmp_{m,i} \Big)^{2},
\end{multline}
where $|\I|$ is the cardinality of set $\I$. The loss function \eqref{eq:loss} characterizes the gap between the Q-value of action $\a_{m}(i)$ given state $\s_{m}(i)$ under the current network parameter vector $\thetae_{m}$ and a target Q-value $\qtmp_{m,i}$ for each experience $i\in\I$ (to be explained in the next paragraph). The minimization of the loss function is accomplished by performing a gradient descent step on the neural network \emph{Eval\_Net$_{m}$}  using  backpropagation (see Section 6 in \cite{goodfellow2016deep}). 

The value  of $\qtmp_{m,i}$ for experience $i\in\I$ is determined based on double-DQN technique \cite{van2016deep}. The double-DQN technique can improve the estimation of the expected long-term cost when compared with the traditional method (e.g., \cite{mnih2015human}). The value of $\qtmp_{m,i}$ for experience $i$ is the sum of the corresponding cost in experience $i$ and a discounted Q-value of the action that is likely to be selected given the next state in experience $i$ under network \emph{Target\_Net$_{m}$}, i.e., 
\begin{equation}\label{eq:q-target1}
	\qtmp_{m,i} =
	c_{m}(i) + \gamma \Qt_{m}(\s_{m}(i+1),\Atmp_i;\thetat_{m}),
\end{equation}
where $\Atmp_i$ is the action with the minimum Q-value given state $\s_{m}(i+1)$ under \emph{Eval\_Net$_{m}$}, i.e.,
\begin{equation}
	\Atmp_i =\arg \min_{\a\in\As} \Qe_{m}(\s_{m}(i+1),\a;\thetae_{m}).
\end{equation}
Intuitively, for experience $i$, the target-Q value $\qtmp_{m,i}$  reveals the expected long-term cost of action $\a_{m}(i)$ given state $\s_{m}(i)$. This is the summation of the actual cost recorded in experience $i$, i.e., $c_{m}(i)$, and the approximate expected long-term future cost based on \emph{Target\_Net$_{m}$}, i.e., $\gamma \Qt_{m}(\s_{m}(i+1),\Atmp_i;\thetat_{m})$.

For every Replace\_Threshold updates, \emph{Target\_Net$_{m}$} will be updated by copying \emph{Eval\_Net$_{m}$}, i.e., $\thetat_{m} = \thetae_{m}$, where  $\text{mod($\cdot$)}$ is the modulo operator (in step 18 in Algorithm \ref{alg:drl2}). The objective of this step is to keep 
the network parameter $\thetat_{m}$ in \emph{Target\_Net$_{m}$} up-to-date, so that it can better approximate the expected long-term cost in the computing of the target Q-values in \eqref{eq:q-target1}.

\section{Performance Evaluation}\label{sec:exp}
In this section, we compare our proposed DRL-based method with  several benchmark methods, including no offloading (denoted by No Offl.), random offloading (denoted by R. Offl.),  PGOA in \cite{yang2018distributed}, and ULOOF in \cite{neto2018uloof}. The PGOA is designed based on the best response algorithm for the potential game, which takes into account the strategic interaction among mobile devices. The ULOOF is designed based on the capacity estimation according to historical observations.  In these simulations, we consider two performance metrics: the ratio of dropped tasks (i.e., the ratio of the number of dropped tasks to the number of total task arrivals) and the average delay (i.e., the average delay of the  tasks whose processing has been completed).  Unless stated otherwise, the basic parameter setting in the simulations are given in Table \ref{table:para}.  In addition,  the  probability  of random exploration $\epsilon$  is set to be gradually decreasing from 1 to 0.01, and the discount factor  $\gamma$ is set to be $0.9$.

In the following, we first show the convergence of the proposed algorithm across episodes. Then, we compare the performance of our proposed algorithm with the benchmark methods under different parameter settings.

\begin{table}[t]
	\centering
	\caption{Parameter settings}\label{table:para}
	\begin{tabular}{ll}
		\hline
		Parameter & Value\\
		\hline
		$M$ & 50 \\ 
		$N$ & 5 \\
		$\Delta$& 0.1 second\\
		$\iotsev_m,m\in\M$ & 2.5 GHz\cite{neto2018uloof}\\
		$\fogsev_n,n\in\N$ & 41.8 GHz\cite{neto2018uloof}\\
		$\trans_{m,n}, m\in\M, n\in\N$& 14 Mbps\cite{cellularspeed}\\
		$\arriv_m(t),m\in\M, t\in\T$& \{2.0, 2.1, \ldots, 5.0\} Mbits \cite{wang2017computation}\\	
		$\density_m,m\in\M$& 0.297 gigacycles per Mbits\cite{wang2017computation}\\
		$\ddl_m,m\in\M$& 10 time slots (i.e., 1 second)\\
		Task arrival probability & 0.3\\
		\hline
	\end{tabular}	
\end{table}

\subsection{Performance and Convergence}
\begin{figure}[t]
	\centering
	\includegraphics[height=3.3cm]{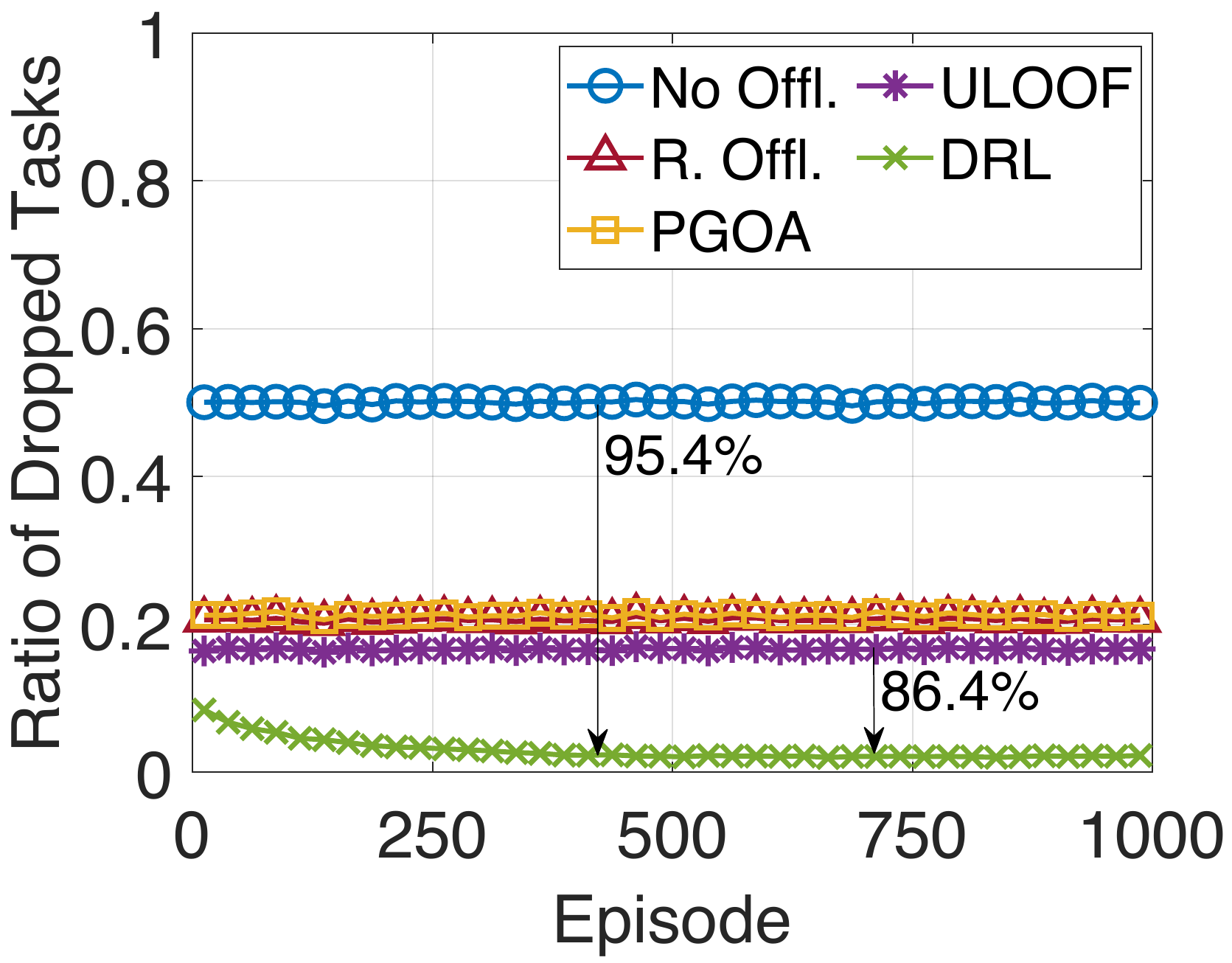}	\includegraphics[height=3.3cm]{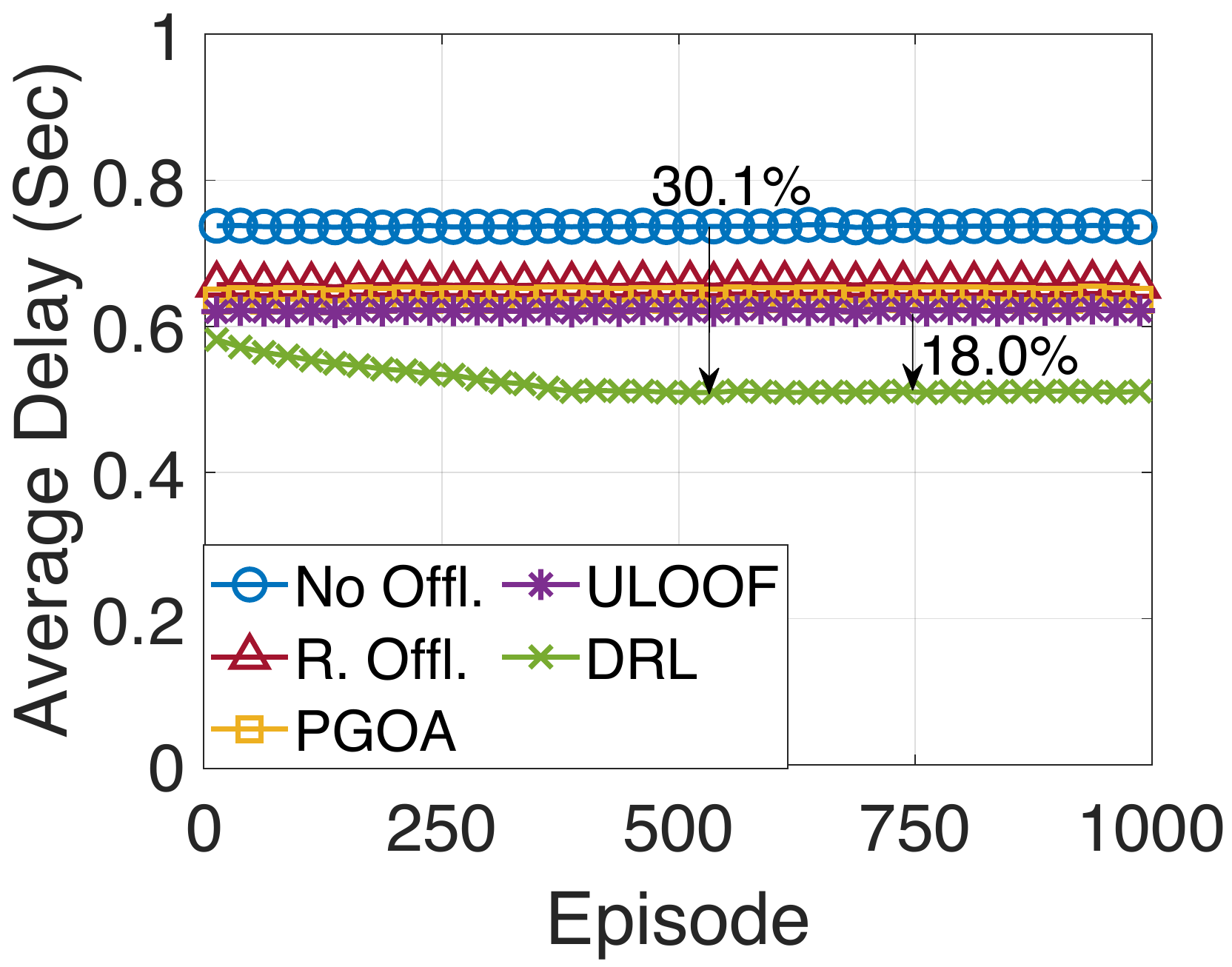}\vspace{-1mm}\\
	\quad(a)\qquad\qquad\qquad\qquad\qquad\qquad(b)\vspace{-2.5mm}\\
	\caption{Performance evaluation across episodes: (a) ratio of dropped tasks; (b) average delay.\vspace{-2mm}}\label{fig:converge}
\end{figure}
Fig. \ref{fig:converge} shows the ratio of dropped tasks and the average delay of the proposed DRL-based algorithm and the benchmark methods across episodes (based on the parameters in Table \ref{table:para}). As shown in  Fig. \ref{fig:converge}, the proposed algorithm converges after around 350 episodes, and it achieves a ratio of dropped tasks of $0.02$ and an average delay of $0.52$ second. This converged performance significantly outperforms those of the benchmark methods.  As shown in the figure, the proposed method reduces the ratio of dropped tasks and the average delay by  $86.4\%-95.4\%$ and $18.0\%-30.1\%$, respectively, when compared with the benchmark methods.

\begin{figure}[t]
	\centering
	\includegraphics[height=3.2cm]{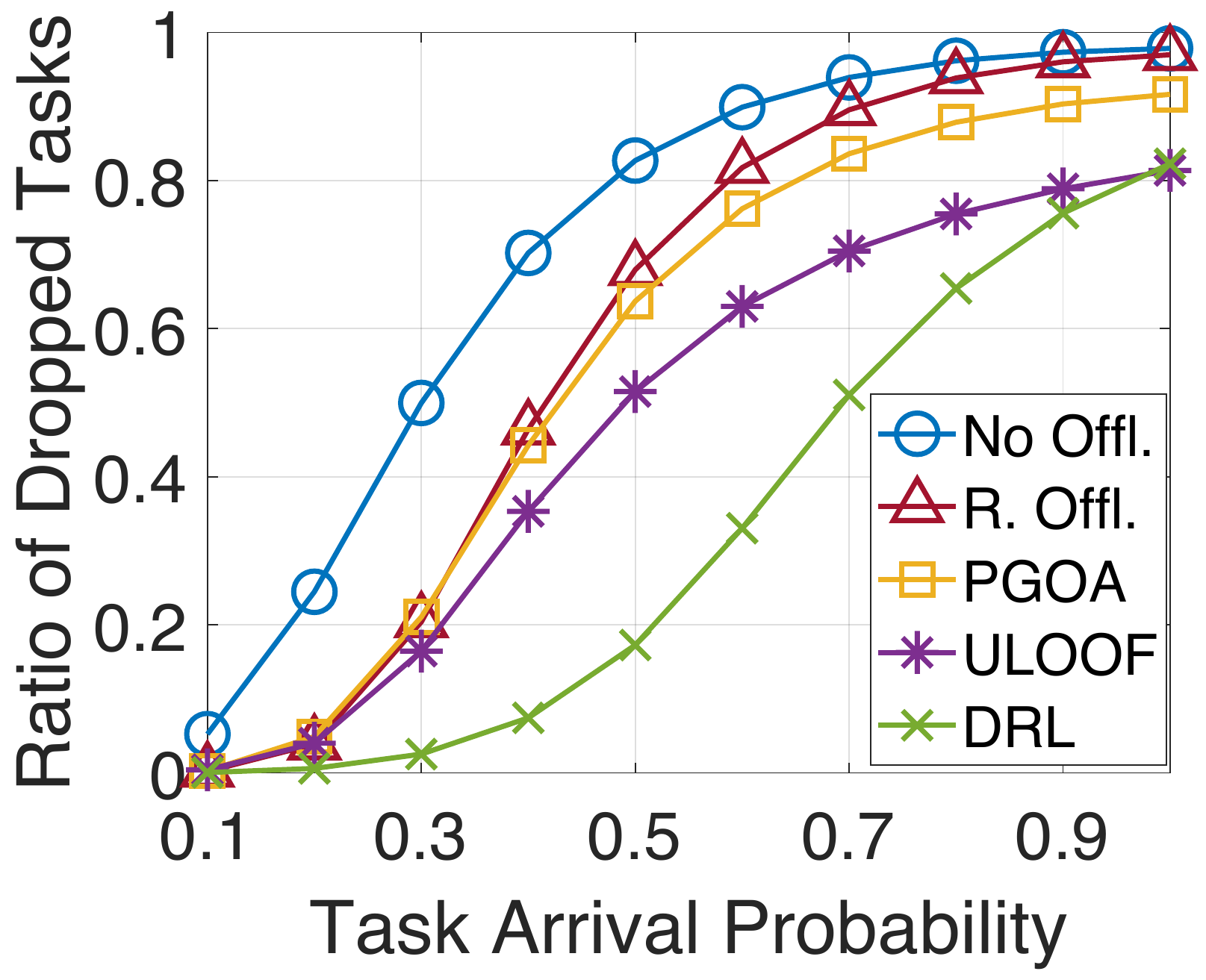}	\includegraphics[height=3.2cm]{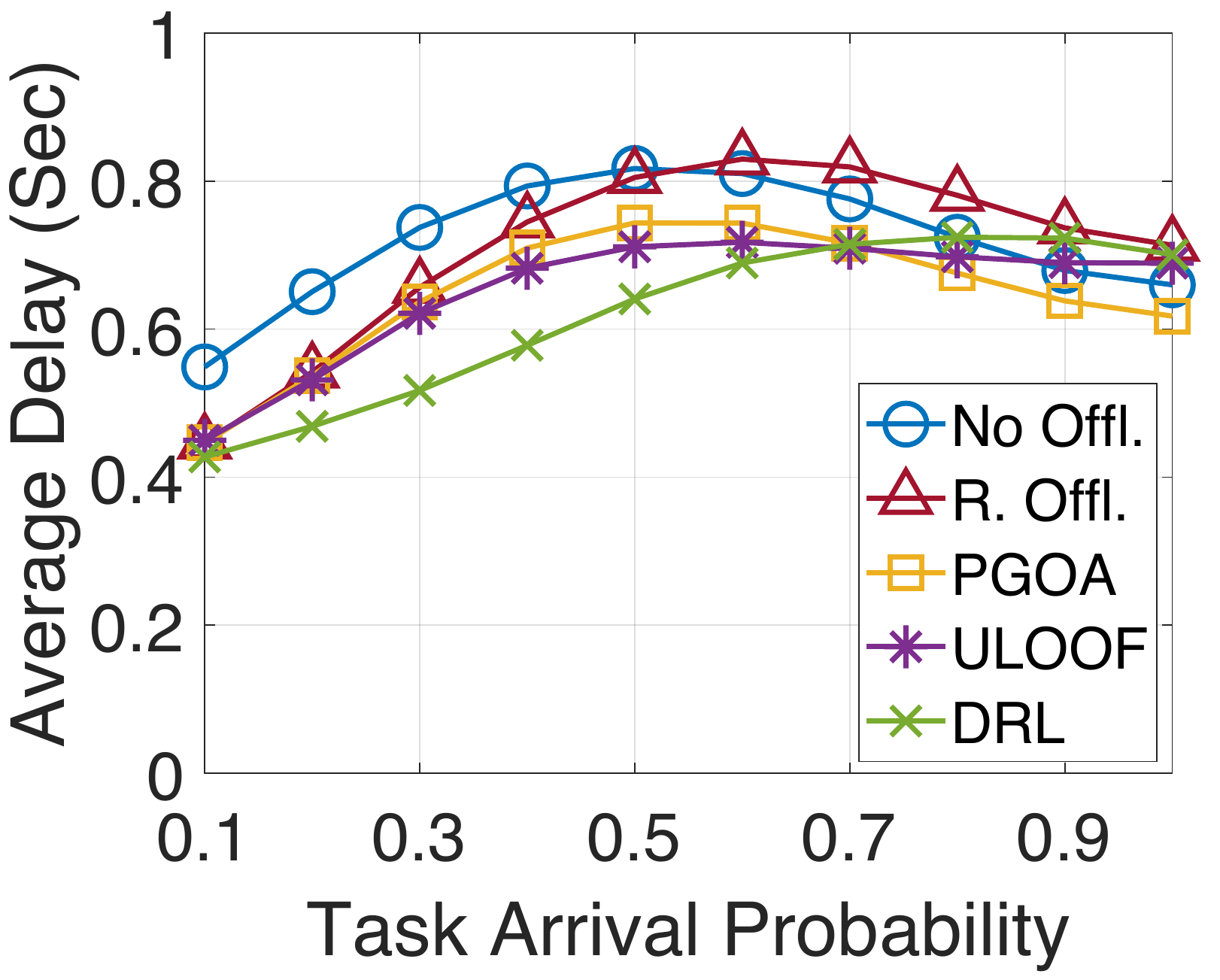}\vspace{-1mm}\\
	\quad(a)\qquad\qquad\qquad\qquad\qquad\qquad(b)\vspace{-2.5mm}\\
	\caption{Performance evaluation under different task arrival probabilities: (a) ratio of dropped tasks; (b) average delay.}\label{fig:arrival-prob}
\end{figure}

\subsection{Task Arrival Probability}
A larger task arrival probability implies a higher load of the system. As shown in Fig. \ref{fig:arrival-prob} (a), as the task arrival probability increases, the proposed DRL-based algorithm can always maintain a lower ratio of dropped tasks when compared with the benchmark methods. Specifically, when the task arrival probability is small (i.e., $0.1$), most of the methods can achieve a ratio of dropped tasks of around zero. As the task arrival probability increases from $0.1$ to $0.5$, the ratio of dropped tasks of the proposed algorithm remains less than $0.2$, while those of the benchmark methods increase to more than $0.5$. In addition, comparing with the benchmark methods, the proposed DRL-based algorithm reduces the ratio of dropped tasks especially when the task arrival probability is moderate (i.e., $0.3-0.8$), where the reduction of the ratio of dropped tasks is at least $13.3\%$. 

In Fig. \ref{fig:arrival-prob} (b),  as the task arrival probability increases from $0.1$ to $0.4$, the average delay of our proposed DRL-based algorithm increases by $26.1\%$, while those of the benchmark methods increase by at least $34.5\%$. This implies that as the load of the system increases, the average delay of the proposed algorithm increases less dramatically than those of the benchmark methods. As the task arrival probability increases to around $0.6$, the average delay of some of the methods decrease, because an increasing number of tasks are dropped and hence are not accounted in the average delay. For the same reason, when the load of the system is high, the proposed algorithm may have a larger average delay  than the other methods, as it has less tasks dropped.

\begin{figure}[t]
	\centering
	\includegraphics[height=3.2cm]{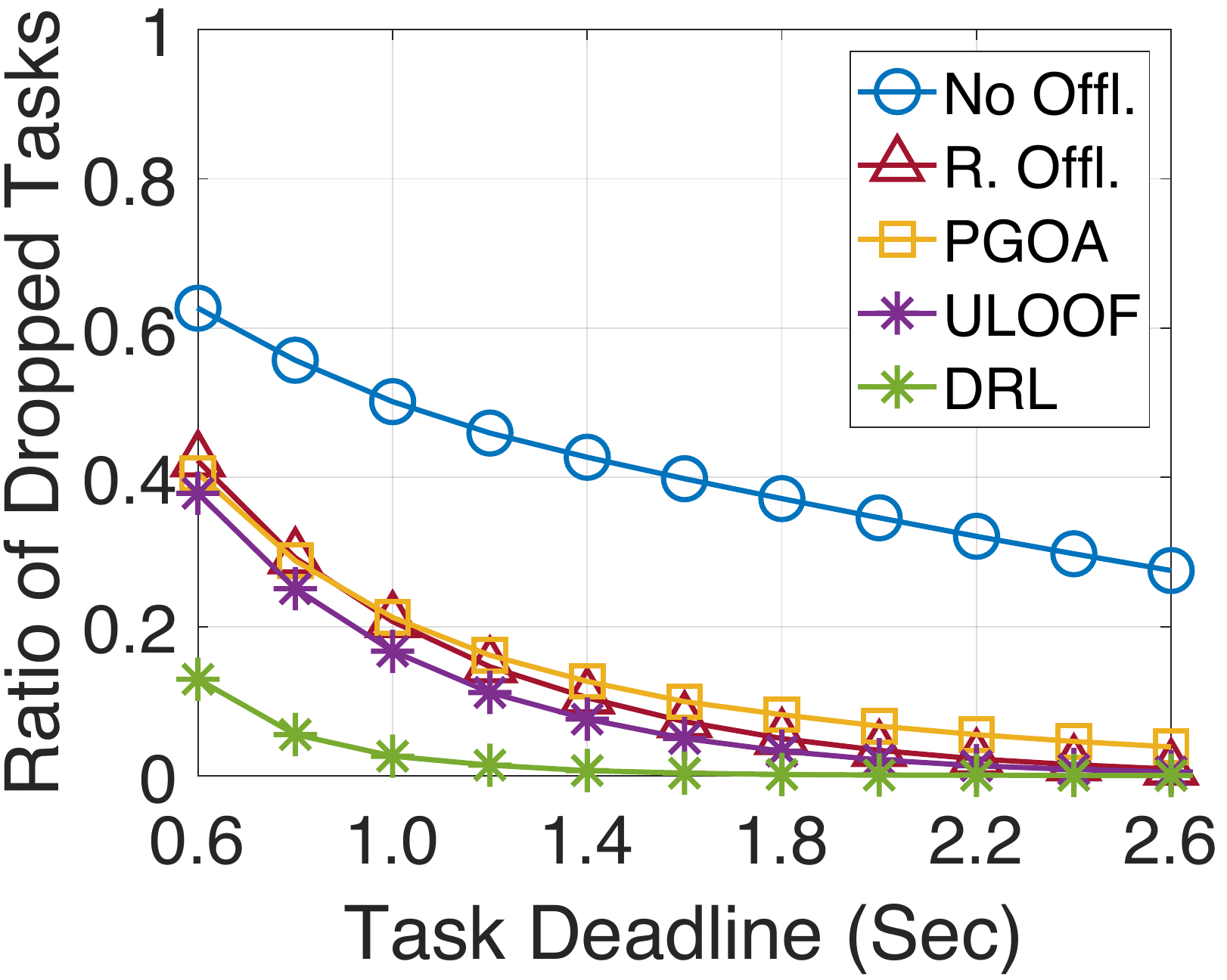}	\includegraphics[height=3.2cm]{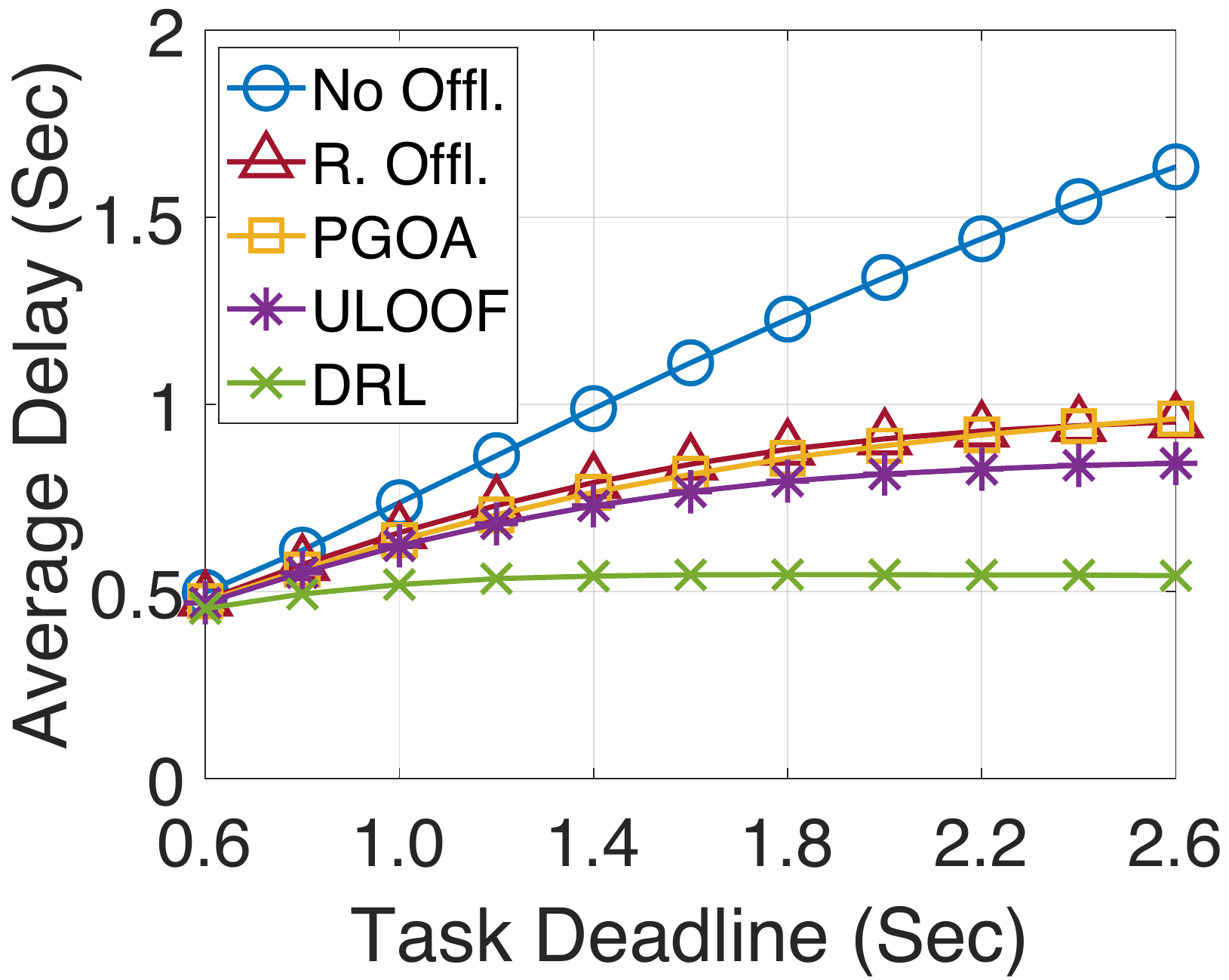}\vspace{-1mm}\\
	\quad(a)\qquad\qquad\qquad\qquad\qquad\qquad(b)\vspace{-2.5mm}\\
	\caption{Performance under different task deadlines: (a) ratio of dropped tasks; (b) average delay.\vspace{-2mm}}\label{fig:ddl}
\end{figure}

\subsection{Task Deadline}
A smaller deadline implies that the tasks are more delay-sensitive. In Fig. \ref{fig:ddl} (a), the proposed algorithm always achieves a lower ratio of dropped tasks than the benchmark methods, especially when the deadline is small. When the task deadline is $0.6$ second, the proposed algorithm reduces the ratio of dropped tasks by $65.8\%-79.3\%$ when compared with the benchmark methods. 
As the deadline increases, the ratio of dropped tasks of each method decreases. With the proposed algorithm, the ratio of dropped tasks is less than $0.01$ when the deadline is larger than $1.4$ seconds. In comparison, the same performance is achieved by ULOOF when the deadline is larger than $2.4$ seconds.

In Fig. \ref{fig:ddl} (b), as the task deadline increases, the average delay of each method increases and gradually converges. This is because when the deadline is larger, the tasks requiring longer processing (and transmission) time can be processed and are accounted in the average delay. When the deadline is large enough, no task is dropped, so further increasing the deadline makes no difference. As shown in Fig. \ref{fig:ddl} (b), the average delay of the proposed algorithm converges (i.e., achieves a marginal increase of less than $0.05$) after the deadline increases to $1.4$ seconds, and the converged average delay is around $0.54$ second. In comparison, the converged average delay of ULOOF is around $0.84$ second, which is $55.6\%$ larger than that of the proposed algorithm, and those of the other methods are larger than $0.96$ second. This implies that when the task deadline is large enough, although each  method can have a ratio of dropped tasks of around zero, the proposed algorithm outperforms the other methods in terms of reducing the average delay.

\begin{figure}[t]
	\centering
	\includegraphics[height=3.2cm]{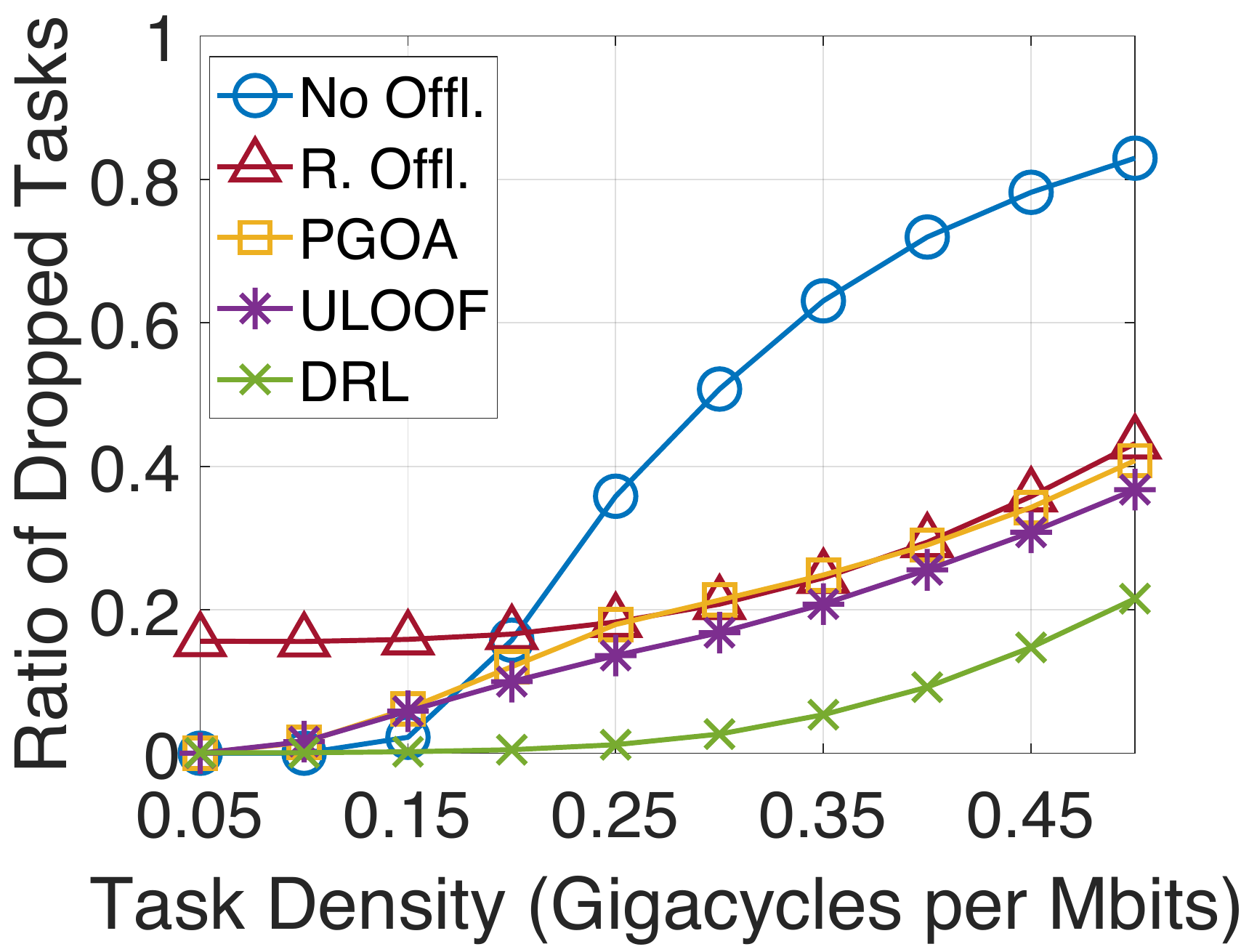}	\includegraphics[height=3.15cm]{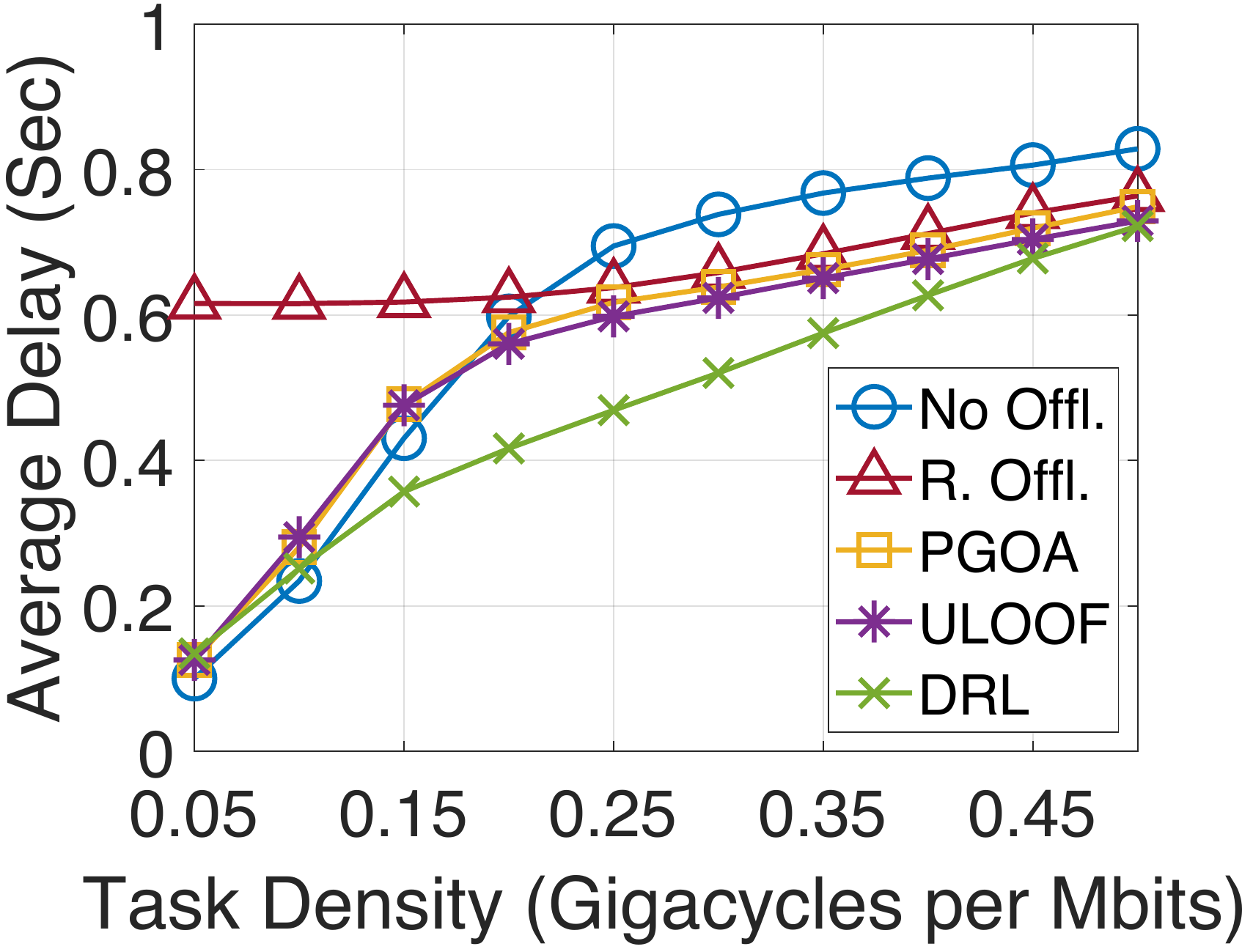}\vspace{-1mm}\\
	\quad(a)\qquad\qquad\qquad\qquad\qquad\qquad(b)\vspace{-2.5mm}\\
	\caption{Performance under different task densities: (a) ratio of dropped tasks; (b) average delay.}\label{fig:density}
\end{figure}

\subsection{Task Density}
A larger task density implies that the computational requirement of each task is larger. As a result, in Fig. \ref{fig:density}, as the task density increases, the ratio of dropped tasks and the average delay of each method increase. On the other hand, when the density is small (e.g., smaller than $0.15$ Gigacycles per Mbits), the transmission delay dominates the processing delay, so no offloading achieves a lower ratio of dropped tasks and a lower average delay than random offloading. When the density is large (e.g., larger than $0.3$ Gigacycles per Mbits), the processing delay dominates the transmission delay, so random offloading achieves a better performance than no offloading.
\begin{figure}[t]
	\centering
	\includegraphics[height=3.2cm]{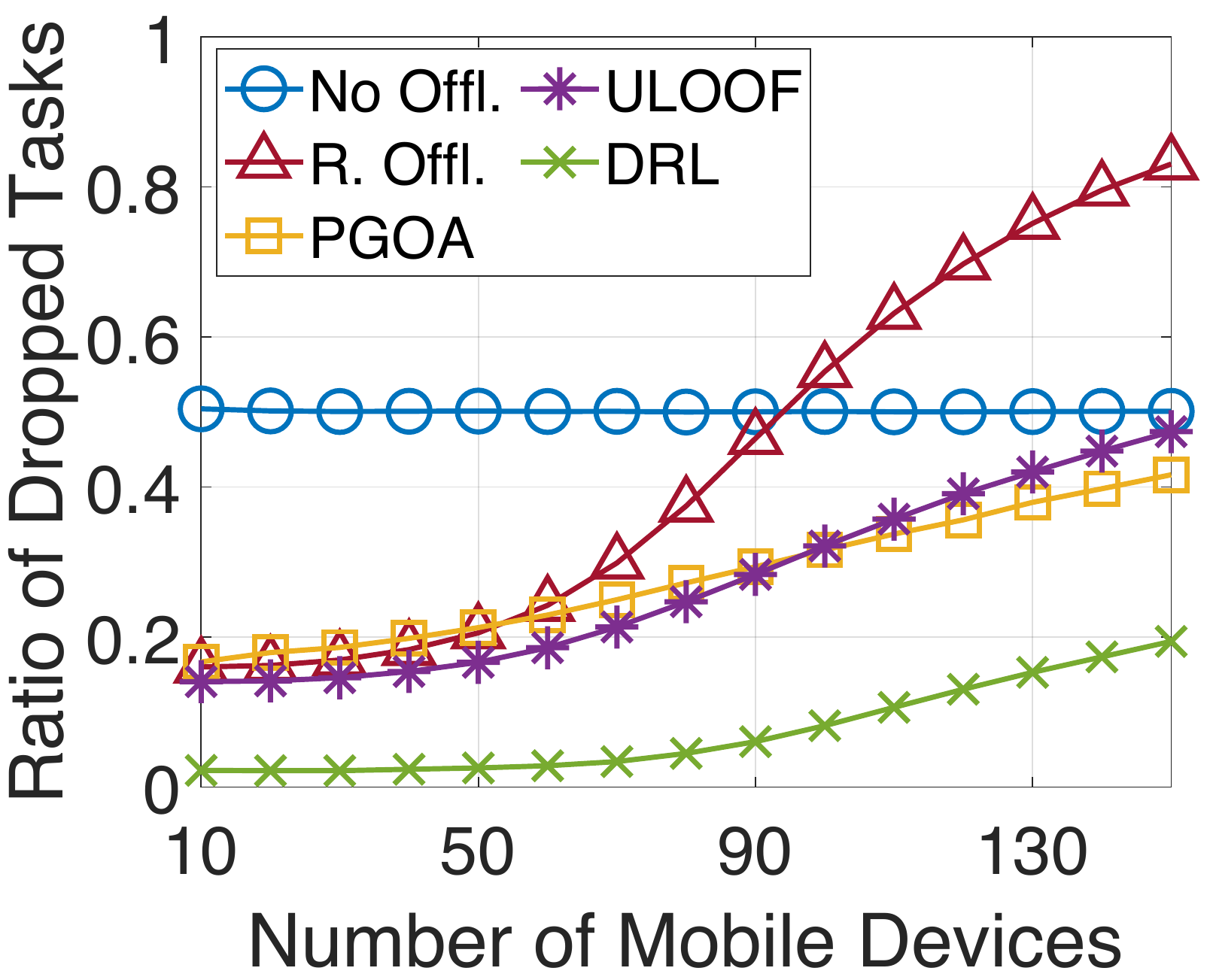}	\includegraphics[height=3.2cm]{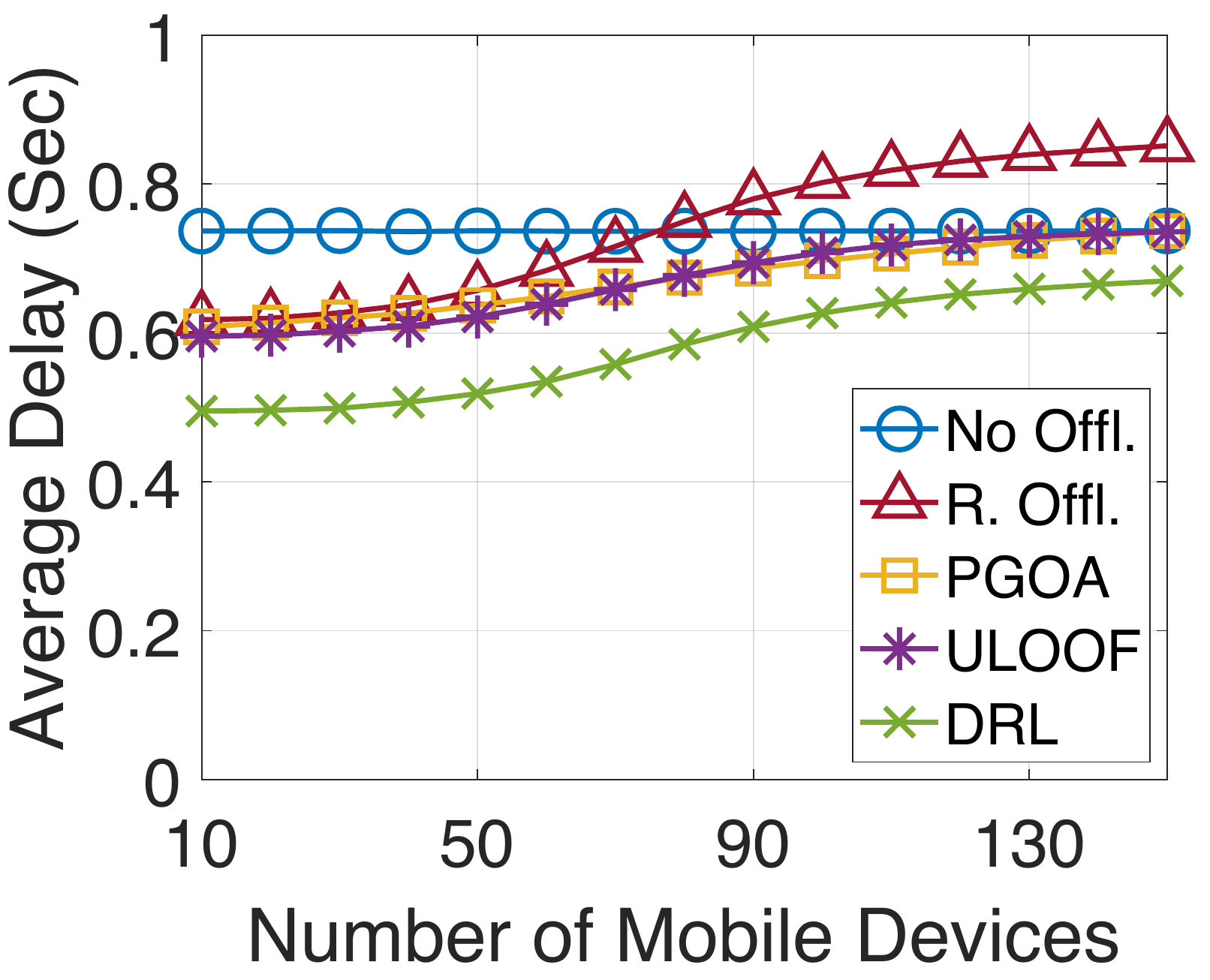}\vspace{-1mm}\\
	\quad(a)\qquad\qquad\qquad\qquad\qquad\qquad(b)\vspace{-2.5mm}\\
	\caption{Performance  under different number of mobile devices: (a) ratio of dropped tasks; (b) average delay.\vspace{-2mm}}\label{fig:iotnum}
\end{figure}

In Fig. \ref{fig:density}, as the task density increases from $0.05$  to $0.25$ Gigacycles per Mbits, the ratio of dropped tasks and the average delay of the proposed algorithm increase less dramatically than those of the benchmark methods. When the density is $0.25$ Gigacycles per Mbits, the proposed algorithm maintains a ratio of dropped tasks of around $0.01$ and an average delay of $0.47$ second. As the task density further increases to $0.5$ Gigacycles per Mbits, although each method achieves a similar average delay, the proposed algorithm can reduce the ratio of dropped tasks by $41.4\%-74.1\%$ when compared with the benchmark methods, because of its proper exploitation of the processing capacities in both the mobile devices and  the edge nodes.

\begin{figure}[t]
	\centering
	\includegraphics[height=3.15cm]{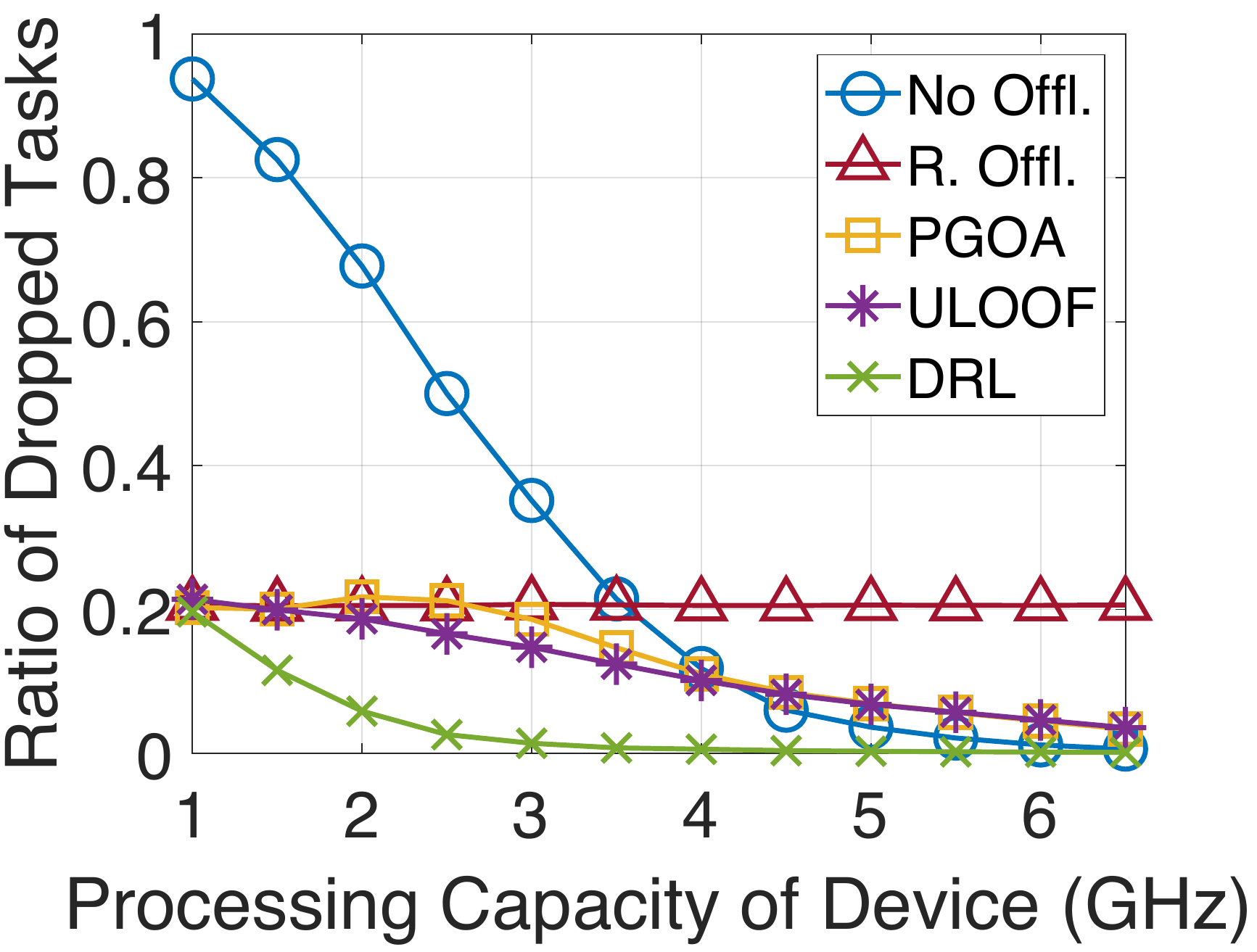}	\includegraphics[height=3.15cm]{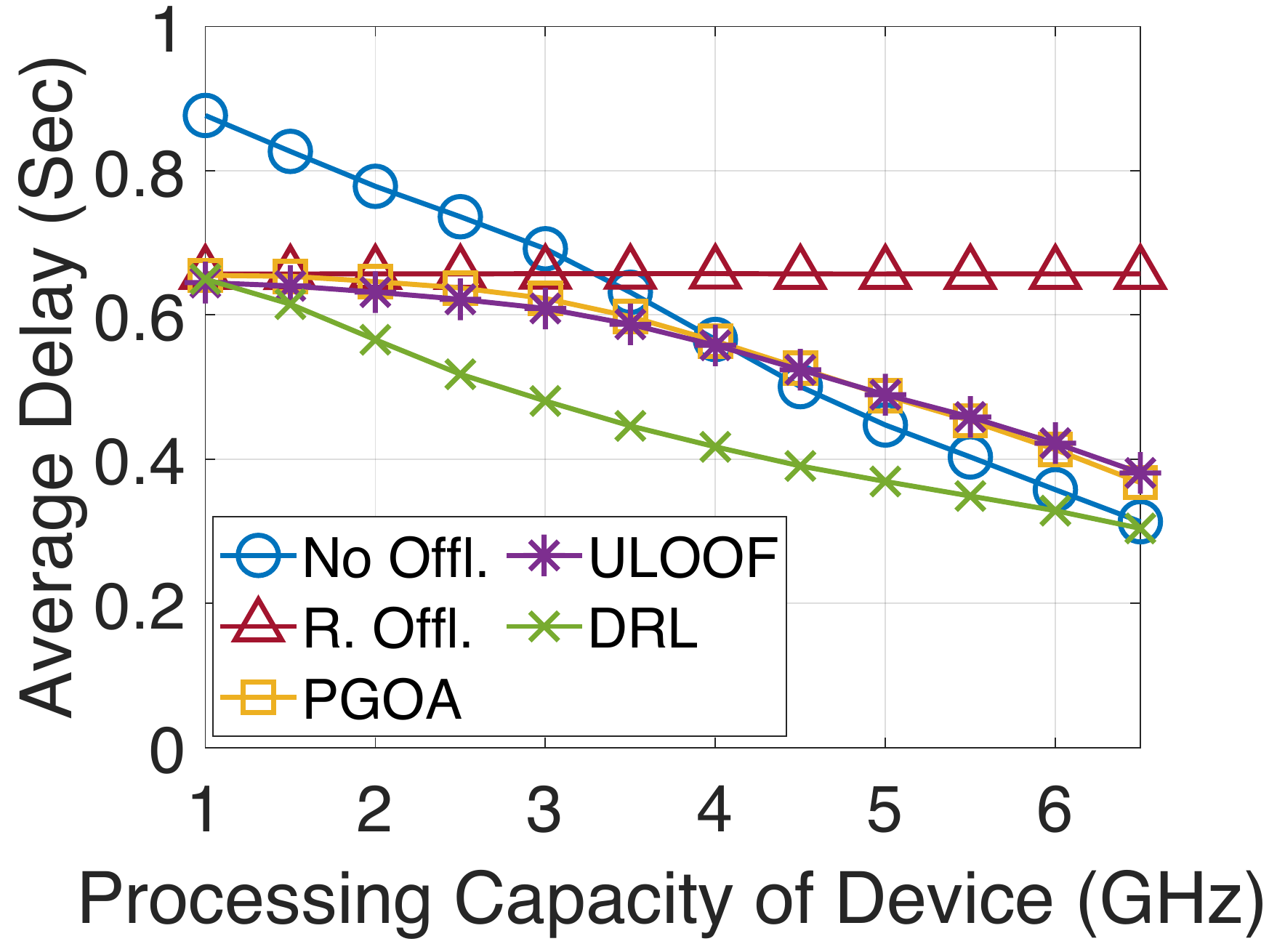}\vspace{-1mm}\\
	\quad(a)\qquad\qquad\qquad\qquad\qquad\qquad(b)\vspace{-2.5mm}\\
	\caption{Performance  under different processing capacities of each mobile device: (a) ratio of dropped tasks; (b) average delay.}\label{fig:iotcap}
\end{figure}

\begin{figure}[t]
	\centering
	\includegraphics[height=3.15cm]{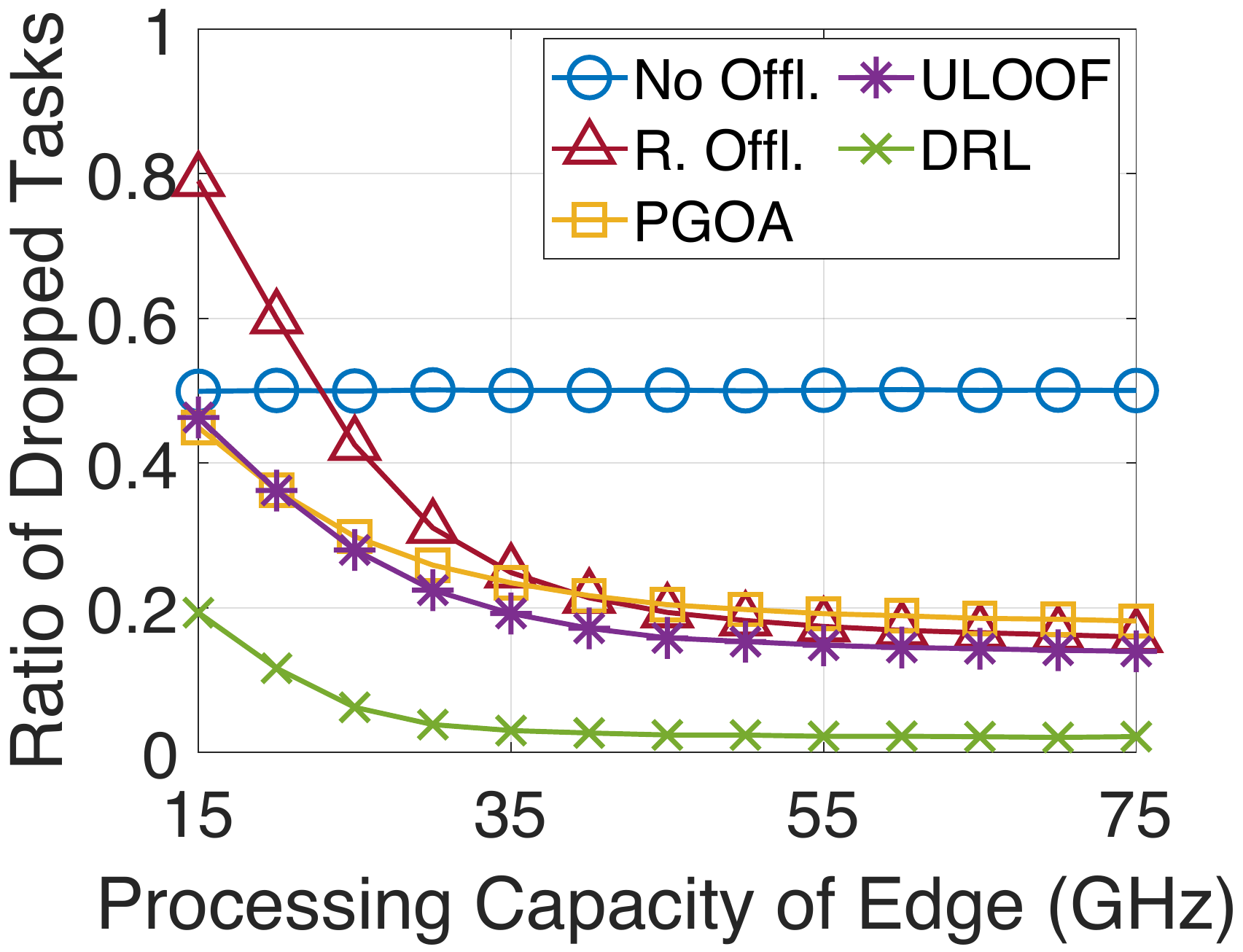}	\includegraphics[height=3.15cm]{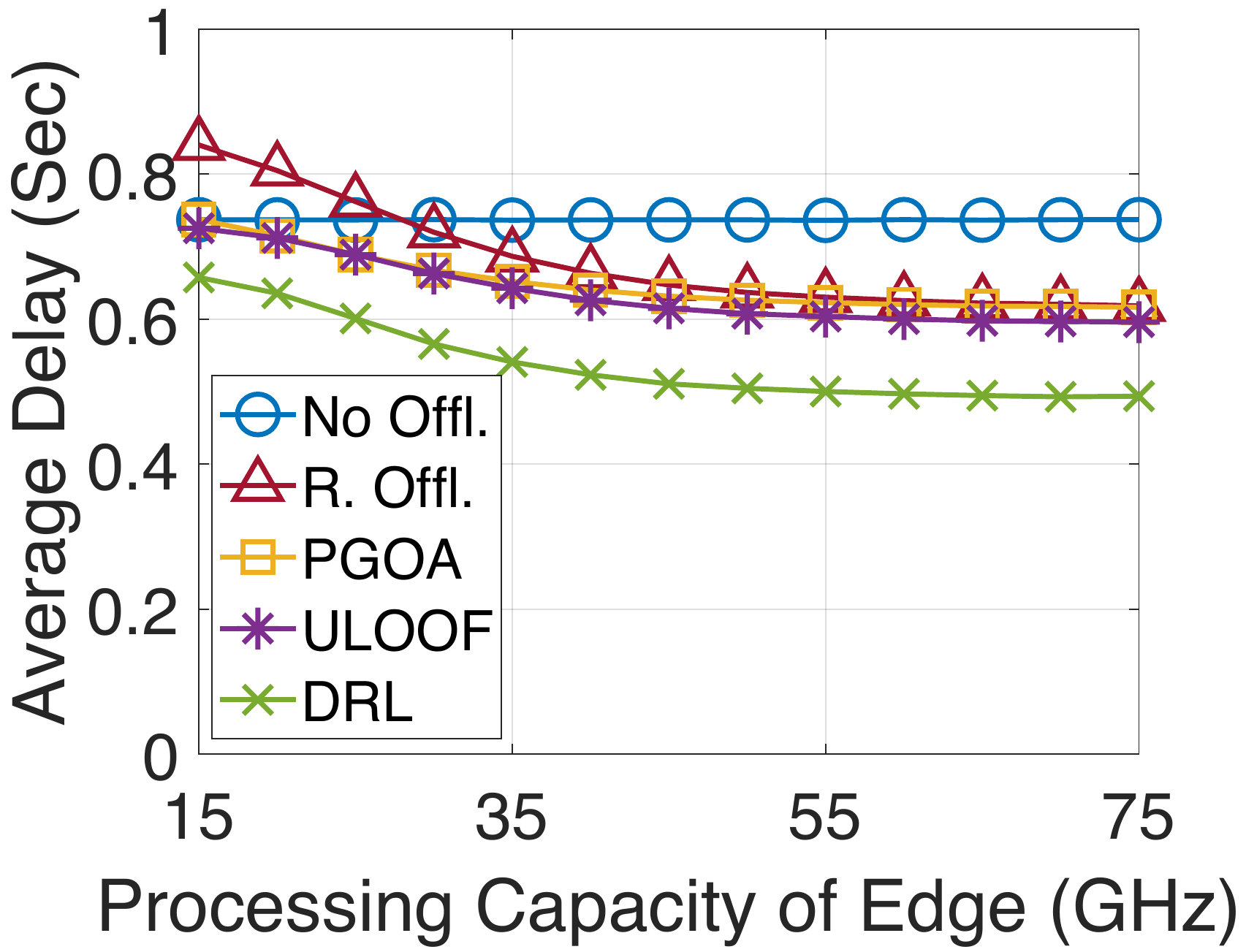}\vspace{-1mm}\\
	\quad(a)\qquad\qquad\qquad\qquad\qquad\qquad(b)\vspace{-2.5mm}\\
	\caption{Performance  under different processing capacities of each edge node: (a) ratio of dropped tasks; (b) average delay.}\label{fig:edgecap}
\end{figure}

\subsection{Number of Mobile Devices}
A larger number of mobile devices implies potentially a higher load at the edge nodes and hence a worse performance for random offloading.  In  Fig. \ref{fig:iotnum} (a), the proposed algorithm achieves a lower ratio of dropped tasks than the other methods, especially when the number of mobile devices is large. This is because the proposed algorithm can effectively address the unknown load dynamics at the edge nodes. When the number of mobile devices increases to $80$, the proposed algorithm maintains a ratio of dropped tasks of less than $0.05$. When it increases to $150$, the proposed algorithm achieves a ratio of dropped tasks of $53.4\%-76.6\%$ less than the benchmark methods.

In Fig. \ref{fig:iotnum} (b), as the number of mobile devices increases, the average delay of each method (except no offloading) increases due to the potentially increasing  load at the edge nodes. Since the proposed algorithm can  effectively deal with the unknown edge load dynamics, when the number of mobile devices increases to $150$, it achieves an average delay of $9.0\%$ lower than those of PGOA and ULOOF.

\subsection{Processing Capacity of Each Mobile Device}
As the processing capacity increases,  the delay of the tasks processed locally decreases. In Fig. \ref{fig:iotcap} (a), as the processing capacity in each device increases, the ratio of dropped tasks of the proposed algorithm decreases more dramatically than those of the benchmark methods. When the processing capacity increases to $3.5$ GHz, the ratio of dropped tasks of the proposed algorithm reduces to $0.007$, which is $93.9\%-96.5\%$ lower than those of the benchmark methods. 

In Fig. \ref{fig:iotcap} (b), as the processing capacity of each device increases, the average delay of the proposed algorithm decreases more dramatically than those of PGOA and ULOOF. When the processing capacity increases to $3.5$ GHz, the average delay of the proposed algorithm is $31.4\%$ and $29.4\%$ lower than those of PGOA and ULOOF, respectively. On the other hand, when the processing capacity of each device is large enough, processing a task locally achieves a strictly lower delay than offloading the task to an edge node due to the transmission time required for offloading, and hence no offloading is optimal. Consequently, as the processing capacity increases, the average delay of the proposed algorithm approaches  the average delay of no offloading.

\subsection{Processing Capacity of Each Edge Node}
With a larger processing capacity of each edge node,  the average delay of the tasks offloaded is smaller. In Fig. \ref{fig:edgecap}, as the processing capacity  increases, the ratio of dropped tasks and the average delay of each method (except no offloading) decrease, because an increasing number of tasks are being offloaded. In addition, those values gradually converge. This is because when the processing capacity of each edge node is large enough,  further increasing the processing capacity does not reduce the  delay of those tasks offloaded due to the limited transmission capacity. 

As shown in Fig. \ref{fig:edgecap}, the proposed algorithm can reduce the ratio of dropped tasks and the average delay when compared with the benchmark methods. The reduction of the ratio of dropped tasks is especially significant when the processing capacity of each edge node is small. When the processing capacity  is $15$ GHz, the proposed algorithm reduces the ratio of dropped tasks by at least $57.0\%$ and reduces the average delay by at least $9.4\%$ when compared with the benchmark methods.  On the other hand, the converged ratio of dropped tasks and the average delay of the proposed algorithm is at least $84.3\%$ and $17.2\%$ less than those of the benchmark methods, respectively. This is because the proposed algorithm can efficiently exploit the processing capacities in the mobile devices and edge nodes as well as the limited transmission capacity for offloading.

In conclusion, comparing with the benchmark methods, our proposed algorithm achieves a lower  ratio of dropped tasks and a lower average delay under the different parameter settings. The reduction of the ratio of dropped tasks is especially significant when the tasks are delay-sensitive or the load levels at the edge nodes are high (i.e., the task density is large, the number of mobile devices is large, or the processing capacity of each edge node is small).

\section{Conclusion}\label{sec:conclusion}
In this work, we studied the computational task offloading problem with  non-divisible and delay-sensitive tasks in the MEC system and designed a distributed offloading algorithm that enables mobile devices to make their offloading decisions in a decentralized manner. The proposed algorithm can address the unknown load level dynamics at the edge nodes, and it can handle the  time-varying system environments (e.g., the arrival of new tasks, the computational requirement of each task). Simulation results showed that when compared with several benchmark methods, our proposed algorithm can reduce the ratio of dropped tasks and average delay. The benefit is especially significant when the tasks are delay-sensitive or the load levels at the edge nodes are high. For future work, it is interesting to enable mobile devices to  learn their optimal offloading policies cooperatively through taking advantage of the trained neural networks of other mobile devices. Through the cooperative learning, the training process of the DRL-based algorithm may be accelerated, and the performance may be improved.

\bibliographystyle{IEEEtran}
\bibliography{surveyset}

\end{document}